\documentclass[a4paper,fleqn,usenatbib]{mnras}
\usepackage{newtxtext,newtxmath}
\usepackage[T1]{fontenc}
\usepackage{ae,aecompl}
\usepackage{graphicx}
\usepackage{amsmath}
\usepackage{csquotes}
\usepackage{subcaption}

\title[Evidence for a dark matter cusp in the tidally disrupting SMC]{Surviving the Waves: evidence for a dark matter cusp in the tidally disrupting Small Magellanic Cloud}
\author[De Leo et al.]{
Michele De Leo,$^{1,2,3}$\thanks{E-mail: micheledl89@gmail.com}
Justin I. Read,$^{3}$
Noelia E. D. No\"{e}l,$^{3}$
Denis Erkal,$^{3}$
Pol Massana,$^{4,3}$
\newauthor{ Ricardo Carrera$^{5}$}
\\
$^{1}$Instituto de Astrof\'{i}sica, Pontificia Universidad Cat\'{o}lica de Chile, Av. Vicu\~{n}a Mackenna 4860, 7820436, Macul, Santiago, Chile\\
$^{2}$Instituto Milenio de Astrof\'{i}sica MAS, Nuncio Monse\~{n}or Sotero Sanz 100, Of. 104, Providencia, Santiago, Chile\\
$^{3}$Department of Physics, University of Surrey, Guildford, GU2 7XH, UK\\
$^{4}$Department of Physics, Montana State University, P.O. Box 173840, Bozeman, MT 59717-3840, USA\\
$^{5}$INAF - Osservatorio Astronomico di Padova, Vicolo dell'Osservatorio 5, 35122 Padova, Italy
}

\date{Accepted XXX. Received YYY; in original form ZZZ}

\pubyear{2024}

\begin{document}
\label{firstpage}
\pagerange{\pageref{firstpage}--\pageref{lastpage}}
\maketitle

\begin{abstract}
We use spectroscopic data for ${\sim}6,000$ Red Giant Branch (RGB) stars in the Small Magellanic Cloud (SMC), together with proper motion data from \textit{Gaia} Early Data Release 3 (EDR3), to build a mass model of the SMC. We test our Jeans mass modelling method (\textsc{binulator}+\textsc{GravSphere}) on mock data for an SMC-like dwarf undergoing severe tidal disruption, showing that we are able to successfully remove tidally unbound interlopers, recovering the dark matter density and stellar velocity anisotropy profiles within our 95\% confidence intervals. We then apply our method to real SMC data, finding that the stars of the cleaned sample are isotropic at all radii (at 95\% confidence) and that the inner dark matter density profile is dense, $\rho_{\rm DM}(150\,{\rm pc}) = 1.58_{-0.58}^{+0.80}\times 10^8 M_{\odot} \rm kpc^{-3} $, consistent with a $\Lambda$ Cold Dark Matter ($\Lambda$CDM) cusp. Our model gives a new estimate of the SMC's total mass within 3\,kpc ($M_{\rm tot} \leq 3\,{\rm kpc})$ of $2.29\pm0.46 \times 10^9 M_{\odot}$. We also derive an astrophysical \textquote{$J$-factor} of $18.99\pm0.16$\, GeV$^2$\,cm$^{-5}$ and a \textquote{$D$-factor} of $18.73\pm0.04$\, GeV$^2$\,cm$^{-5}$, making the SMC a promising target for dark matter annihilation and decay searches. Finally, we combine our findings with literature measurements to test models in which dark matter is \textquote{heated up} by baryonic effects. We find good qualitative agreement with the Di Cintio et al. 2014 model but we deviate from the Lazar et al. 2020 model at high $M_*/M_{200} > 10^{-2}$. We provide a new, analytic, density profile that reproduces dark matter heating behaviour over the range $10^{-4} < M_*/M_{200} < 10^{-1}$.
\end{abstract}


\section{Introduction}\label{intro}

One of the long-standing problems of the prevailing $\Lambda$ Cold dark matter ($\Lambda$CDM) cosmological model is the discrepancy between the observed constant density \textquote{cores} of gas rich dwarf galaxies ($\rho_{DM}(150\,{\rm pc}) \sim$ constant $\sim 5 \times 10^7$\,M$_\odot$\,kpc$^{-3}$; e.g. \citealt{1994Natur.370..629M,1994ApJ...427L...1F,2017MNRAS.467.2019R}) and the dense \textquote{cusps} predicted by pure dark matter structure formation simulations ($\rho_{DM}(150\,{\rm pc}) > 10^8$\,M$_\odot$\,kpc$^{-3}$; e.g. \citealt{1991ApJ...378..496D,1996ApJ...462..563N,1997ApJ...490..493N}). Numerous solutions to this so-called \textquote{cusp-core problem} have been proposed, falling into three main categories. The first class of solution proposes new dark matter models, such as Self Interacting dark matter \citep{2000PhRvL..84.3760S}, Warm dark matter \citep[e.g.][]{2000PhRvD..62f3511H,2001ApJ...556...93B,2001ApJ...559..516A}, or \textquote{Wave-like} dark matter \citep[e.g.][]{2014NatPh..10..496S}. The second class challenges the interpretation of the data in some cases, for example the existence of systematic errors due to the typically assumed spherical symmetry and circular gas motions \citep[e.g.][]{2016MNRAS.462.3628R,2018MNRAS.474.1398G,2019MNRAS.482..821O}. The third class proposes that \textquote{baryonic effects}, like repeated gas cooling and blowout through the starburst cycle, can kinematically \textquote{heat} the dark matter pushing it out of the centres of dwarf galaxies \citep[e.g.][]{1996MNRAS.283L..72N,2002MNRAS.333..299G,2001ApJ...560..636E,2005MNRAS.356..107R,2008Sci...319..174M,2012MNRAS.421.3464P,2014MNRAS.437..415D,2014MNRAS.441.2986D,2014Natur.506..171P,2021MNRAS.504.3509O}. This third class of solution has been gaining traction due to it making a number of testable predictions that are now supported by a host of observational data. dark matter heating models predict that star formation should be bursty, with a peak-to-trough burst amplitude of $\sim$10, a burst duration shorter than the local dynamical time and a kinematically \textquote{hot} stellar disc \citep[e.g.][]{2013MNRAS.429.3068T,2017MNRAS.466...88S}. The same models predict that stars should slowly migrate outwards \citep{2005MNRAS.356..107R}, yielding an age gradient \citep[e.g.][]{2016ApJ...820..131E} and that cusp-core transformations need to take many dynamical times, meaning that dwarf galaxies with truncated star formation should be more cuspy than those with extended star formation \citep[e.g.][]{2014MNRAS.437..415D,2016MNRAS.459.2573R}. All of these predictions have been borne out by data so far \citep[e.g.][]{2014MNRAS.441.2717K,2012ApJ...750...33L,2019ApJ...881...71E,2012AJ....143...47Z,2019MNRAS.484.1401R,2022NatAs...6..647C}.

However, a key forecast of dark matter heating models has only recently been tested. Following \citet{2012ApJ...759L..42P}, \citet{2014MNRAS.437..415D} parameterise the amount of cusp-core transformation a dwarf galaxy undergoes by its stellar-to-halo mass ratio, $M_*/M_{200}$. This works to leading order\footnote{In practice, $M_*/M_{200}$ is not fully sufficient on its own as it does not capture information about the size of the dark matter core (which is typically of order the half stellar mass radius, $R_{1/2}$; \citealt{2015MNRAS.454.2092O,2016MNRAS.459.2573R}), the burstiness of the star formation that actually took place, nor the impact of potential fluctuations driven by gas/stellar clumps and/or minor mergers \citep[e.g.][]{2001ApJ...560..636E,2021MNRAS.504.3509O}. Nonetheless, $M_*/M_{200}$ does appear to correlate well with the presence/absence of a core for most simulated dwarfs in a $\Lambda$CDM cosmology \citep[e.g.][]{2014MNRAS.437..415D}.} because $M_*$ is proportional to the total integrated supernova energy available to unbind the dark matter cusp, while $M_{200}$ represents the potential well depth and, therefore, how much energy is required. \citet{2014MNRAS.437..415D} predict cusped dwarfs for $M_*/M_{200} \lesssim 5 \times 10^{-4}$, cored dwarfs for $5 \times 10^{-4} \lesssim M_*/M_{200} \lesssim 5 \times 10^{-2}$, and cusped dwarfs again for $M_*/M_{200} \gtrsim 10^{-2}$, with this latter owing to the potential well depth winning over the energy available to unbind the cusp.\footnote{This prediction may need to be revisited, however, if Active Galactic Nuclei in dwarfs provide an additional source of significant potential fluctuations \citep[e.g.][]{2013MNRAS.432.1947M}.} 
\citet{2019MNRAS.484.1401R} measured the inner dark matter densities of 16 nearby dwarfs with  $10^{-4} \lesssim M_*/M_{200} \lesssim 5 \times 10^{-3}$, finding excellent qualitative agreement with \citet{2014MNRAS.437..415D}. In a similar study, \citet{2022A&A...658A..76B} probed $10^{-3} \lesssim M_*/M_{200} \lesssim 3 \times 10^{-2}$, finding results consistent with \citet{2019MNRAS.484.1401R} where they overlap, and favouring a return cuspy galaxies at higher $M_*/M_{200}$, as predicted by \citet{2014MNRAS.437..415D}. However, \citet{2022A&A...658A..76B} base their study on dwarfs at a redshift $z=1$ that are not necessarily comparable with the local sample from \citet{2019MNRAS.484.1401R}. In this context, the Small Magellanic Cloud (SMC), with $M_*/M_{200} \sim 7 \times 10^{-3}$, and at a distance of  $\sim62\pm1.28$ kpc \citep{2020ApJ...904...13G} from us poses a unique opportunity to test dark matter heating models at a higher $M_*/M_{200}$ than previously probed for nearby dwarfs.
The main challenge to using a standard equilibrium mass modelling method in this galaxy is the overwhelming evidence showing that the outskirts of the SMC are in fact tidally disrupted (e.g. \citealt{2008MNRAS.386..826E,2011ApJ...737...29O,2013ApJ...768..109N,2014MNRAS.442.1897R,2014MNRAS.442.1663D,2015MNRAS.452.4222N,2017MNRAS.471.4571C};\citealt{2018ApJ...864...55Z,2019ApJ...874...78Z,2020MNRAS.498.1034M,2020MNRAS.495...98D,2021ApJ...910...36Z,2021MNRAS.502.2859N}). The hypothesis of heavy tidal disruption is also supported by the observations of distance-tracer populations such as classical Cepheids \citep[i.e.][]{2016AcA....66..149J,2016ApJ...816...49S,2017MNRAS.472..808R} and RR Lyrae \citep[i.e.][]{2017AcA....67....1J,2018MNRAS.473.3131M} that show a long line-of-sight depth for the SMC.

Fortunately, there is no direct observational evidence that the tidal disruption extends to the inner regions of the SMC. It is thus possible to reconcile the observations of extended disruption and long line-of-sight depth previously mentioned with equilibrium mass modelling by hypothesising that the SMC is composed of a bound remnant surrounded by an extended field of tidal debris. This bound remnant is not intended as a classical stellar bulge, but rather as the ensemble of all matter (dark matter, stars and gas) still bound by the SMC self-gravity. Due to the SMC being a dark matter dominated system, we expect this bound remnant to be dominated by the dark matter distribution and exhibit a roughly spherical shape. This would be the case even if the SMC possessed gaseous and stellar discs in its past as such structures don't survive tidal stirring and evolve into spherical distributions \citep[i.e.][and references therein]{2013ApJ...764L..29K}. The key to successfully model the galaxy’s remnant then resides in the removal of the debris along the line of sight to the centre of the SMC that can contaminate the inner stellar kinematics \citep[i.e][]{2007MNRAS.378..353K}.

In this paper, we combine the unprecedented kinematic sample of ${\sim}6000$ SMC stars from \citet{2020MNRAS.495...98D}, which includes line of sight velocities and proper motions, with the Jeans modeling code \textsc{GravSphere}\footnote{Available here: https://github.com/justinread/gravsphere} \citep{2017MNRAS.471.4541R,2018MNRAS.481..860R,2020MNRAS.498..144G,2021MNRAS.505.5686C} to produce a new mass model of the SMC. We assume that the SMC's tidal disruption is not complete, such that the central bound region of the galaxy can be modelled assuming pseudo-dynamic equilibrium. We use a new binning module for \textsc{GravSphere} called the \textsc{binulator} \citep{2021MNRAS.505.5686C} to successfully remove contaminating, tidally unbound, stars along the line of sight to the SMC. To test our method, we apply the \textsc{binulator} and \textsc{GravSphere} to mock data for a severely disrupting SMC, showing that even in this extreme case, we are able to correctly infer the stellar velocity anisotropy and inner dark matter density within our 95\% confidence intervals. We then apply \textsc{binulator} and \textsc{GravSphere} to the real SMC data to constrain the dark matter mass profile of the SMC, its pre-infall mass and to test dark matter heating models. We also determine whether the SMC is a promising target for dark matter annihilation and/or decay searches \citep[e.g.][]{2016PhRvD..93f2004C}.

This paper is organised as follows. In \S\ref{data}, we present the data used for the mass model. In \S\ref{modeling}, we discuss the \textsc{GravSphere}+\textsc{binulator} mass modelling method. In \S\ref{res}, we present our results which we discuss in \S\ref{discussion}. Finally, we present our conclusions in \S\ref{conclusions}.

\section{Data}\label{data}
In this section, we describe both the observational (\S\ref{obs}) and simulation data (\S\ref{sims}) that we use to generate the mock data for this study.

\subsection{Observational data}\label{obs}
We used spectroscopic data of RGB stars in the SMC area from the catalogues presented in \citet{2020MNRAS.495...98D} and \citet{2014MNRAS.442.1663D}, the \textit{Gaia} DR2 \citet{2018A&A...616A...1G} and EDR3 \citet{2021A&A...649A...1G} catalogues, and a photometric selection of RGB stars from the Survey of the MAgellanic Stellar History \citep[SMASH,][]{2021AJ....161...74N}, cross-matched with \textit{Gaia} EDR3.

\subsubsection{Radial velocities}\label{rv}
\begin{figure}
	\includegraphics[width=\columnwidth]{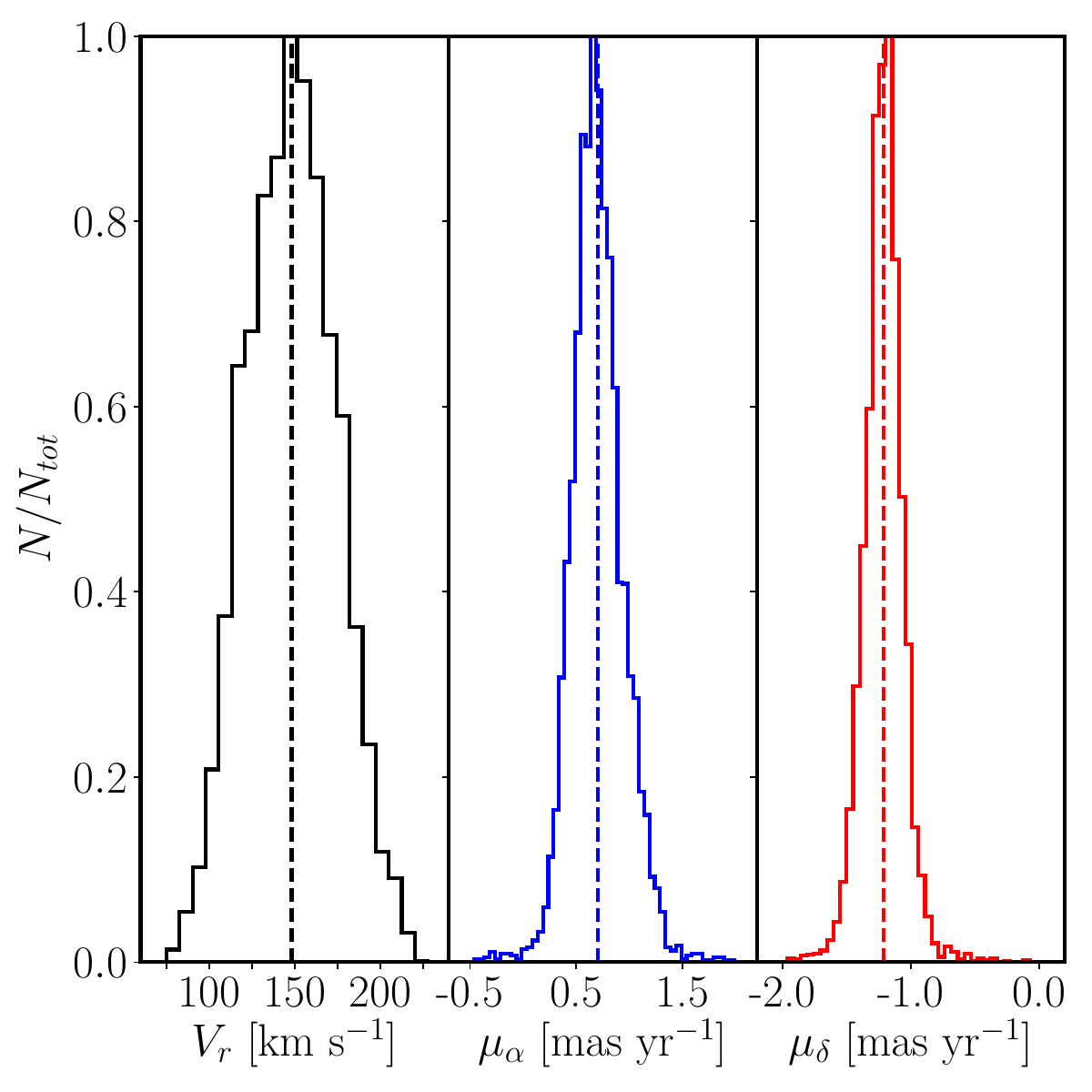}
    \caption{Kinematic distributions of our sample stars. Each panel shows an histogram that represents the distribution of one of the velocity components; the perpendicular dashed lines represent the mean values. The velocity components shown are the radial velocity $V_r$ (in black in the \textit{left panel}), the proper motion $\mu_{\alpha}$ (in blue in the \textit{middle panel}), and the proper motion $\mu_{\delta}$ (in red in the \textit{right panel}).}
    \label{fig:kinematics_distributions}
\end{figure}

We used the radial velocity determinations from the \textquote{extended} sample presented in \citet{2020MNRAS.495...98D} which also includes SMC RGB stars from \cite{2014MNRAS.442.1663D}. For the full details of the analysis that led to the radial velocity determinations see \citet{2020MNRAS.495...98D}. Briefly, the raw spectra acquired with the 2dF+AAOmega instrument at the AAT were processed with the 2dfdr tool\footnote{See https://www.aao.gov.au/science/software/2dfdr} \citep{2010PASA...27...91S} and proprietary software to reduce them, remove sky contamination, subtract the solar reflex motion and finally derive radial velocities through cross-correlation with a grid of synthetic spectra (details of the grid in \citealt{2018A&A...618A..25A}). This sample includes $\sim6000$ RGB stars which are confirmed SMC members (i.e. with radial velocities $V_r$ between 70 and 230 km s$^{-1}$). The distribution of radial velocities can be seen on the left panel of Fig.~\ref{fig:kinematics_distributions} where the large velocity dispersion of the system \citep[c.f.][]{1993MNRAS.261..873H,2006AJ....131.2514H} is clearly appreciated.

\subsubsection{Proper motions}\label{pms}
We cross-matched the radial velocity sample presented above with the \textit{Gaia} EDR3 catalogue. For discussions on the systematics of \textit{Gaia} see \citet{2018A&A...616A...2L}, the recommendations from L. Lindegren\footnote{IAU 30 GA \textit{Gaia} 2 astrometry talk, available in extended version at
https://www.cosmos.esa.int/web/gaia/dr2-known-issues.} and \citet{2021A&A...649A...2L}. The total error budget for the proper motions in \textit{Gaia} is as follows:
\begin{equation}
    \sigma_{tot} = \sqrt{k^2 \sigma_i^2 + \sigma_s^2}
\end{equation}
where $k$ is a factor accounting for the underestimation of the observational uncertainties, $\sigma_i$ is the measured uncertainty for the \textit{i}-th star and $\sigma_s$ is the systematic error. The main difference with the data presented in 
\citet{2020MNRAS.495...98D} (which used proper motions from \textit{Gaia} DR2) is in the lower observational uncertainties and systematics of \textit{Gaia} EDR3 which translated into an improvement of about 65\% in the systematic error $\sigma_s$ and total uncertainties which are on average 47\% and 35\% smaller, respectively for $\mu_{\alpha}\cos{\delta}$ and $\mu_{\delta}$. Throughout the paper we will refer to the proper motion $\mu_{\alpha}\cos{\delta}$ simply as $\mu_{\alpha}$. As done in \citet{2020MNRAS.495...98D}, we corrected the proper motion measurements to account for the solar reflex motion and for geometric effects (systematic contraction/expansion with respect to the centre due to the motion along the line of sight) following \citet{2006A&A...445..513V} and \citet{2018MNRAS.481.2125B}:
\begin{equation}\label{geometric_eff}
    \mu_r=-6.1363\times10^{-5}V_{los}\frac{R}{d}
\end{equation}
where $\mu_r$ is the correction in mas\,yr$^{-1}$, $V_{los}$ is the velocity along the line of sight in km\,s$^{-1}$, $R$ is the distance to the centre of the SMC in arcmin and $d$ is the distance between the SMC and the Sun in kpc. The distributions of proper motions can be seen in blue ($\mu_{\alpha}$) and in red ($\mu_{\delta}$) respectively in the middle and right panels of Fig.~\ref{fig:kinematics_distributions}. Both distributions show long tails and the distribution of $\mu_{\alpha}$ (the blue histogram in the middle panel) shows larger dispersion and asymmetry favouring proper motions higher than the mean.

\subsubsection{Stellar surface density}\label{surf_den}
\begin{figure}
    \centering
    \includegraphics[width=\columnwidth]{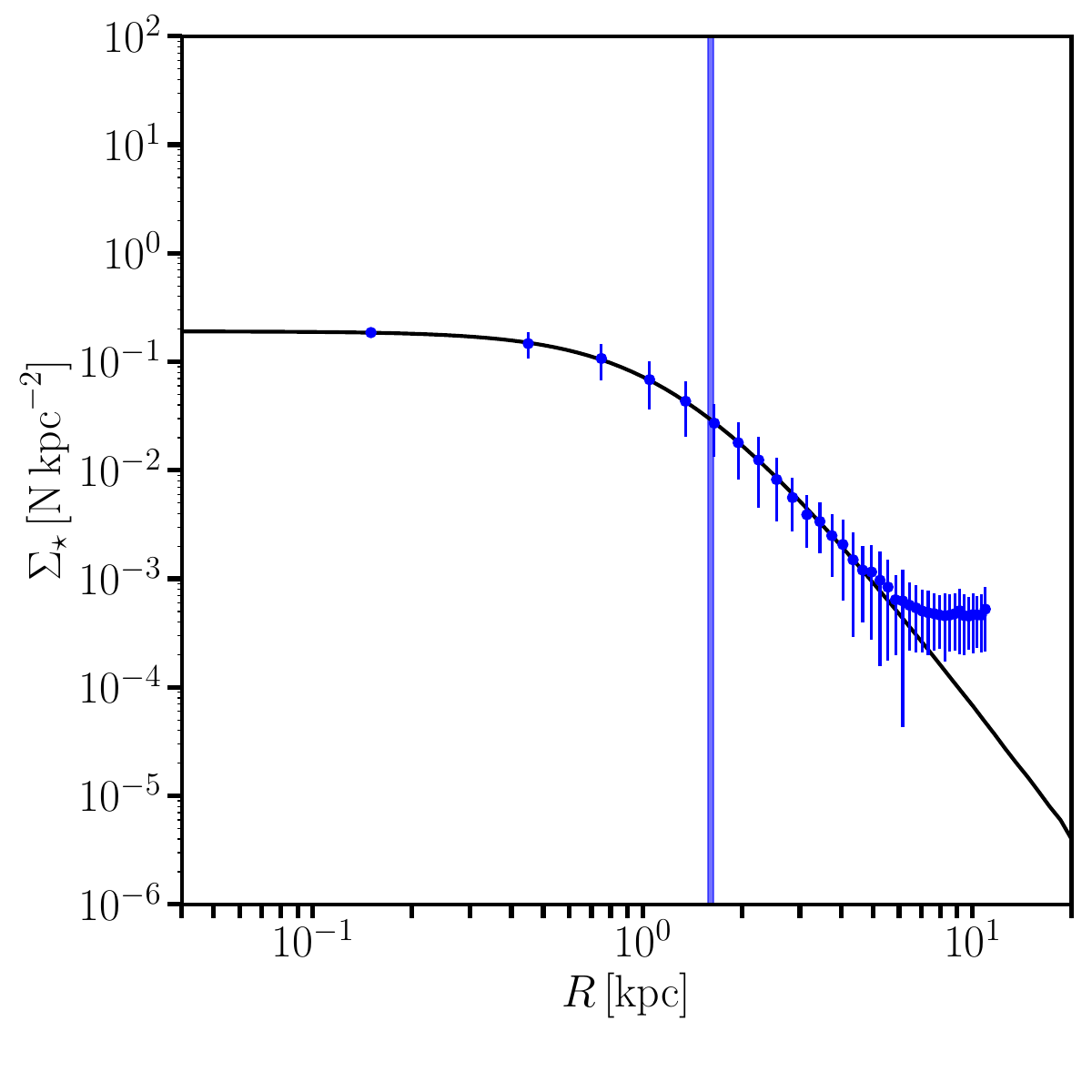}
    \caption{Surface number density profile, $\Sigma_*(r)$, of the SMASH sample of RGB stars cross-matched with Gaia EDR3 \citep{2020MNRAS.498.1034M}. The black line is the best fit model from \textsc{GravSphere}, the blue points with 68\% confidence intervals are the profile computed from real data, and the fainter blue vertical line is the half-light radius computed by \textsc{GravSphere}.
    }
    \label{fig:smash_prof}
\end{figure}

The kinematic sample presented above (radial velocities plus proper motions) was too small and incomplete to provide a reliable estimation of the stellar surface density. To have a more complete sampling of the inner regions of the SMC we derived the stellar surface density from a selection of RGB stars from SMASH, crossmatched with \textit{Gaia} EDR3 \citep[a sample from][]{2020MNRAS.498.1034M}. Using the deep photometry of SMASH and proper motions and parallaxes from \textit{Gaia} EDR3 for foreground decontamination, we produced an accurate density profile of upper RGB bona-fide SMC candidates out to $\sim$11$^{\circ}$ from the centre of the galaxy (that for this sample is at $\alpha_c$, $\delta_c$ = 13.16$^{\circ}$, -72.80$^{\circ}$). The density is computed first counting the number of stars in each HEALPix\footnote{http://healpix.sourceforge.net} \citep{2005ApJ...622..759G,Zonca2019} pixel (nside=512), then dividing the SMC projected surface in equal-radius (0\fdg3) annuli centred on ($\alpha_c$, $\delta_c$) and averaging the number of stars over the pixels included in each given annulus (taking into account that all pixels have equal area). Given the large number of RGB stars present in the SMASH sample, we were able to compute statistical uncertainties on the average values by taking the standard deviations in each annulus. This provided an accurate stellar surface density profile based on the same type of stars of our kinematics tracers and useful over large radii (see Figure \ref{fig:smash_prof}).

\subsection{Simulation data}\label{sims}
In order to test our analysis method and mass model, we used two SMC-analogues taken from the suite of simulations presented in \citet{2020MNRAS.495...98D}. One is a \textquote{bound} SMC that has undergone little tidal stripping. The other is a \textquote{heavily disrupted} SMC that is close to full dissolution. This latter is closest to the real SMC data though is likely more extreme. In particular, it starts out close to the SMC's {\it current} inner velocity dispersion before tidal stripping and shocking. As such, its final inner velocity dispersion is substantially colder than the true SMC data. This is nonetheless perfectly acceptable for testing our methodology.

\begin{figure}
     \centering
     \hspace{-9mm}
     \includegraphics[width=1.1\columnwidth]{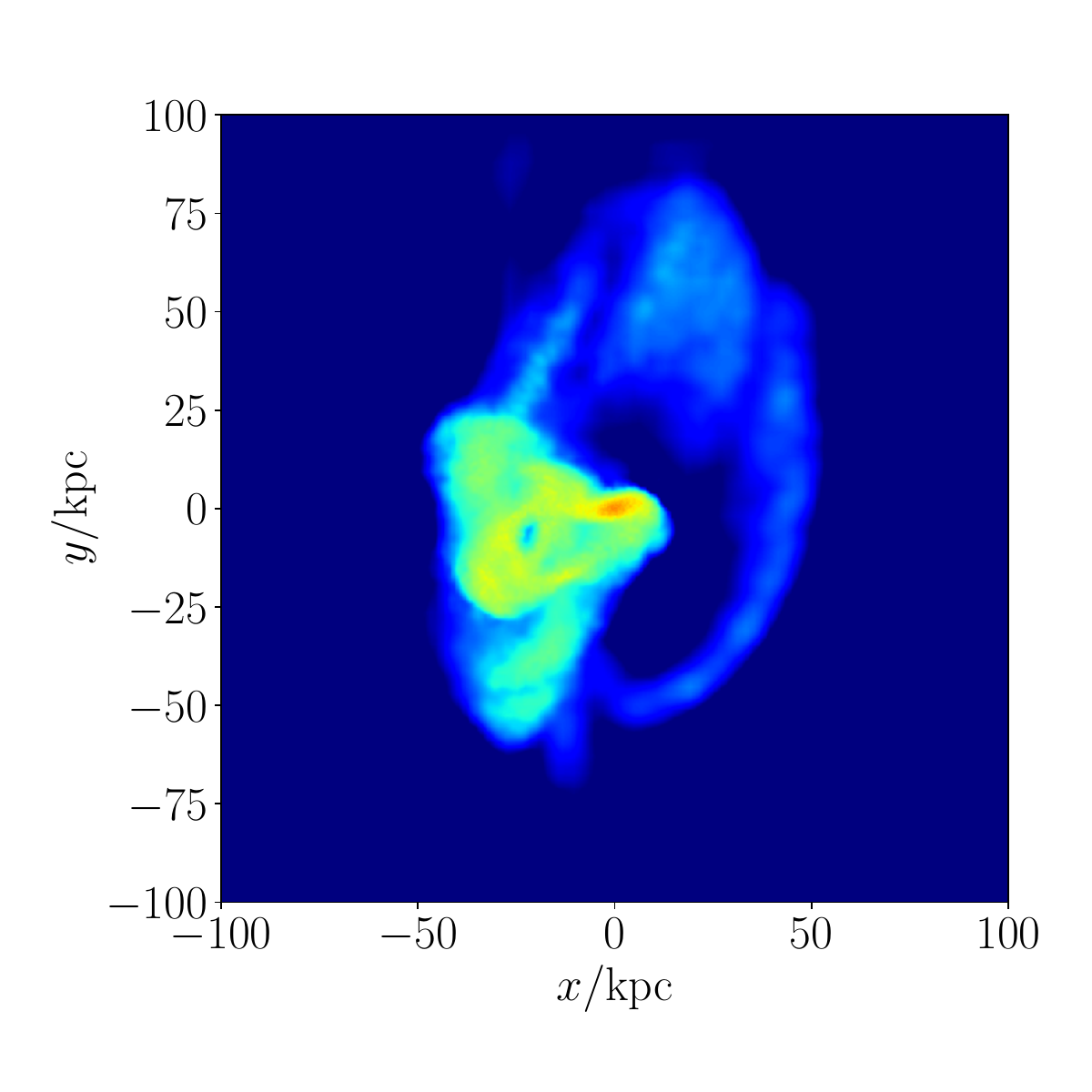}
     \vspace{-9mm}
     \caption{Density map of the heavily disrupted simulation  \citep{2020MNRAS.495...98D}) projected on the $XY$-plane of the simulation phase space and centred on its photometric centre. Blue pixels have no particles in them. The density increases from cyan to light green, yellow and orange for the densest regions in the centre. The cyan and light green plumes and arcs trace the past orbit of the simulated galaxy and are composed of particles making up tidal debris around the central bound remnant of the galaxy.}
     \label{fig:denmap_surfden}
\end{figure}

The simulation setup is described in detail in \citet{2020MNRAS.495...98D}. Here, we briefly summarise the key points. We ran a grid of $N$-body simulations using \textsc{gadget-3} \citep[an updated version of \textsc{gadget-2};][]{2005MNRAS.364.1105S} with a live SMC ($10^5$ particles, a total mass of $M_{\rm SMC} = 10^9 M_{\odot}$, and two different scale radii, as outlined below) modelled as a Plummer sphere \citep{1911MNRAS..71..460P} disrupting around a $1.5\times10^{11}\,{\rm M}_\odot$ LMC \citep{2019MNRAS.487.2685E} in the presence of the Milky Way (modelled using the \texttt{MWPotential2014} model from \citealt{2015ApJS..216...29B}). As specified in the original paper, the initial mass used for the SMC is based on its present-day dynamical mass \citep[i.e.,][]{2006AJ....131.2514H}, substantially lower than its likely peak mass before infall ($\sim 5\times10^{10}\,{\rm M}_{\odot}$; \citealt{2019MNRAS.487.5799R}). This is motivated by the fact that the simulations were not meant to faithfully model the entire disruption history of the SMC but only its final phases, once the SMC had lost the majority of its dark matter \citep{2016ApJ...833..109S}. The two simulations selected for the analysis in this work are at the opposite ends of the spectrum of explored scale radii, the most bound (0.8\,kpc initial scale radius) and the most disrupted (1.5\,kpc initial scale radius) SMC-analogues. The former bound SMC is used to test that the method works properly and is able to recover the true density profile in the absence of disrupting influences; the latter heavily disrupted SMC is the closest analogue to our observed case and likely, in fact, more extreme (see Figure \ref{fig:denmap_surfden} for a visual representation).

In \citet{2020MNRAS.495...98D} we have shown how the heavily disrupted simulation qualitatively reproduces the kinematic trends of the dispersion of the velocity distributions of the real SMC (visible here by comparing the top left panel of Fig. \ref{fig:full_2s_escvel_comparison}, for the heavily disrupted simulation, and the left panel of Fig. \ref{fig:binulator_power}, for the reals SMC). Figure \ref{fig:denmap_surfden} highlights how the debris field of tidally disrupted simulation particles (low density cyan and blue-ish) is widely spread, covering tens of kpc in the sky. To further prove the similarity of the heavily disrupted simulation with the real SMC morphology, we computed the projected-on-the-sky ellipticity of both (ignoring the distance information of the simulation, unavailable for the real SMC). Following \citet{2016MNRAS.456..870B}, we obtained a projected ellipticity of 0.962 using the tracers within 1\,kpc of the centre for the simulation data at the final snapshot, which compares well to the 0.943 at the same radius obtained for the real SMC data with the same technique.

\begin{figure}
\centering
\includegraphics[width=\columnwidth]{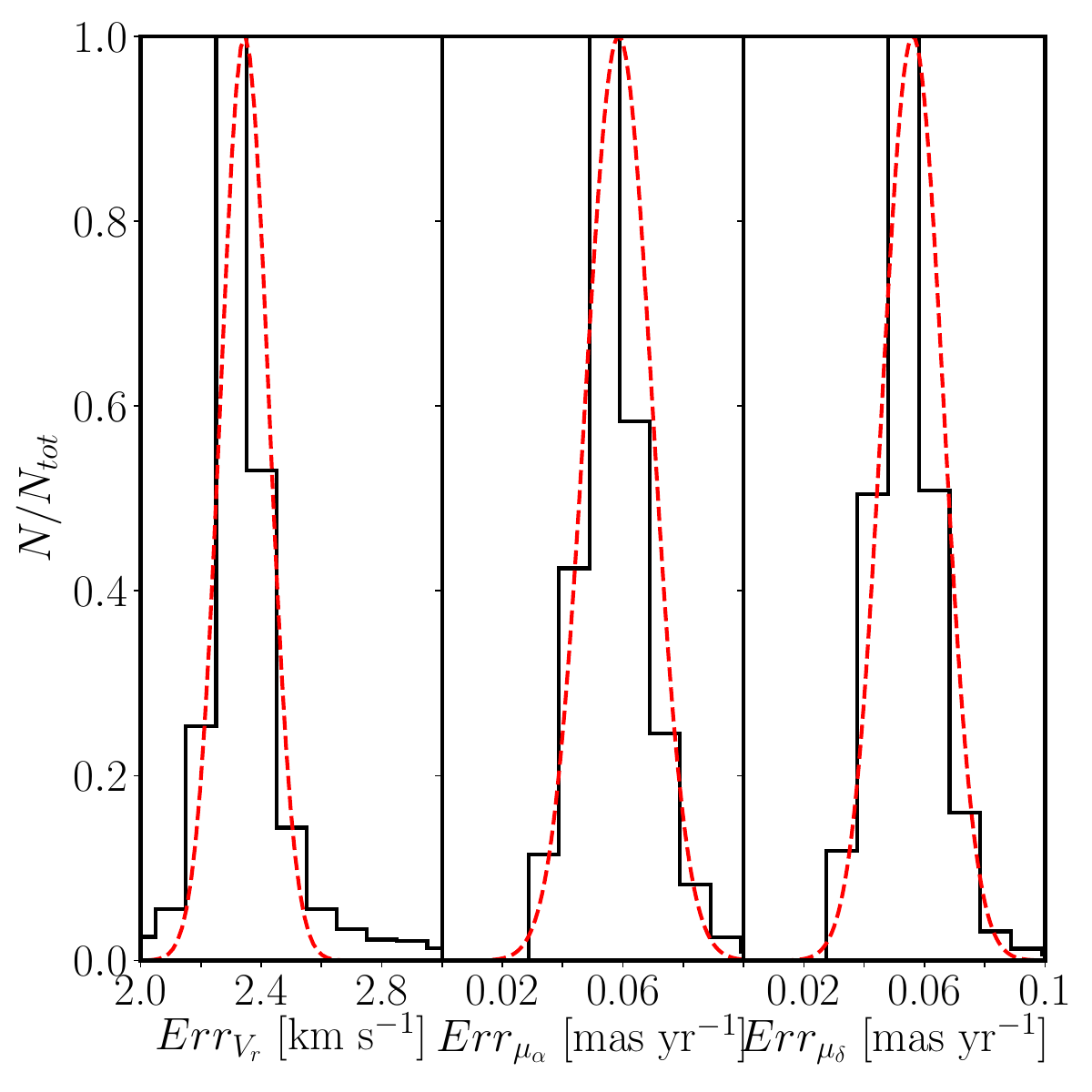}
\caption{Distributions of the observed errors for the real SMC data (black histograms) and Gaussians from which the errors for the simulated mock SMC-analogues are sampled (red dash-dotted lines). Each panel shows a different variable: $V_r$ (\textit{left panel}), $\mu_{\alpha}$ (\textit{middle panel}), and $\mu_{\delta}$ (\textit{right panel}). \label{fig:error_dists_sims}}
\end{figure}

Both the observed and the simulation data were treated with the same analysis pipeline and mass modeling tool. This procedure included adding errors for the simulation data resembling the observational ones. The errors for the simulation data were sampled from Gaussians for each observable ($V_r$, $\mu_{\alpha}$ or $\mu_{\delta}$). The results of this procedure can be seen in Fig.~\ref{fig:error_dists_sims}, where in each panel the black histograms are the observed errors (left panel for $\rm{Err_{V_r}}$, middle panel for $\rm{Err_{\mu_{\alpha}}}$ and right panel for $\rm{Err_{\mu_{\delta}}}$) and the red dot-dashed lines are the Gaussians sampled to generate the simulation errors.

Regarding the errors for the simulation data, it is important to keep in mind that, as pointed out in \citet{2020MNRAS.495...98D}, the simulated SMC analogues are less massive than the real SMC, leading to kinematically cooler velocity dispersion profiles (see also the discussion, above). Thus, when drawing our errors from the real SMC observations, we overestimate the fractional uncertainty on the simulated dispersion profiles. This means that our mock data present a worst-case scenario for testing our analysis pipeline.

\section{Mass Modelling}\label{modeling}

\subsection{\textsc{GravSphere}}\label{gravsphere}

Our mass modelling technique is based on solving the spherical Jeans equation \citep[e.g.][]{1922MNRAS..82..122J,1980MNRAS.190..873B,1987gady.book.....B}:

\begin{equation}\label{jeans_sph}
    \frac{1}{\nu}\frac{\partial }{\partial r}(\nu \sigma_r^2)+\frac{2\beta(r)\sigma_r^2}{r}=-\frac{GM(<r)}{r^2}
\end{equation}
where $\nu(r)$ is the kinematic tracer density; $\sigma_r^2=\langle v_r^2\rangle-{\langle v_r\rangle}^2$ is the velocity dispersion of the tracers; $G=6.67398\times10^{-11}$ $\rm{m^3 {kg}^{-1} s^{-2}}$ is Newton's gravitational constant; $M(<r)$ is the total mass inside spherical radius $r$; and $\beta(r)=1-\sigma_t^2/\sigma_r^2$ is the velocity anisotropy, where $\sigma_t$ is the tangential velocity dispersion. This equation is valid under the assumptions of no rotation (which we excluded for the SMC RGB population in  \citealt{2020MNRAS.495...98D}), dynamical equilibrium, and spherical symmetry. We will test these latter two assumptions with mock data in \S\ref{sims}.

We solve equation \ref{jeans_sph} using the \textsc{GravSphere} code with the aim of recovering the total cumulative mass (stars and dark matter), the dark matter density profile $\rho_{DM}(r)$ and the stellar velocity anisotropy profile $\beta(r)$ of the object being studied. The full methodology is described and tested in detail in \citet{2017MNRAS.471.4541R,2018MNRAS.481..860R,2020MNRAS.498..144G,2021MNRAS.501..978R,2021MNRAS.505.5686C}. The code is publicly available\footnote{\href{https://github.com/justinread/gravsphere}{https://github.com/justinread/gravsphere}.} and has already been used to model a wide range of nearby spherical stellar systems \citep[e.g.][]{2018MNRAS.481..860R,2019MNRAS.484.1401R,2021MNRAS.505.5686C,2021arXiv211209374Z}. Here, we briefly summarise the main points. 

Eq.~\ref{jeans_sph} is integrated along the line of sight to obtain expressions for the line of sight, radial and tangential velocity dispersions, as a function of projected distance, $R$ (e.g. \citealt{1982MNRAS.200..361B,1994MNRAS.270..271V,2005MNRAS.363..705M}):

\begin{equation}\label{jeans_true}
    \sigma_{LOS}^2=\frac{2}{\Sigma_*(R)}\int_R^\infty \bigg(1-\beta(r)\frac{R^2}{r^2}\bigg)\frac{\nu(r)\sigma_r^2(r)r}{\sqrt{r^2-R^2}}\rm{d}r
\end{equation}
\begin{equation}\label{posradial}
    \sigma_{POSr}^2=\frac{2}{\Sigma_*(R)}\int_R^{\infty}\bigg(1-\beta(r)+\beta(r)\frac{R^2}{r^2}\bigg)\frac{\nu(r)\sigma_r^2(r)r}{\sqrt{r^2-R^2}}\rm{d}r
\end{equation}
\begin{equation}\label{postang}
    \sigma_{POSt}^2=\frac{2}{\Sigma_*(R)}\int_R^{\infty}(1-\beta(r))\frac{\nu(r)\sigma_r^2(r)r}{\sqrt{r^2-R^2}}\rm{d}r
\end{equation}
where $\sigma_{LOS}$, $\sigma_{POSr}$ and $\sigma_{POSt}$ are the tracers' line of sight, projected radial and projected tangential velocity dispersions; $\Sigma_*(R)$ is the tracers' surface density at the projected radius $R$; and 
\begin{equation}\label{disp_mid}
    \sigma_r^2(r)=\frac{1}{\nu(r)g(r)}\int_r^{\infty}\frac{GM(\widetilde{r})\nu(\widetilde{r})}{{\widetilde{r}}^2}g(\widetilde{r})\rm{d}\widetilde{r}
\end{equation}
with:
\begin{equation}\label{gierre}
    g(r)={\rm exp} \left(2\int \frac{\beta(r)}{r} {\rm d}r \right)
\end{equation}

\textsc{GravSphere} has the possibility of additionally using two Virial Shape Parameters (VSP1 and VSP2) that constrain the global kurtosis of the stellar distribution, allowing the degeneracy between the cumulative mass and the velocity anisotropy, present if only line-of-sight velocity data are available, to be broken \citep[e.g.][]{1990AJ.....99.1548M,2014MNRAS.441.1584R,2017MNRAS.471.4541R}. However, for the SMC we have excellent constraints on $\sigma_{LOS}$, $\sigma_{POSr}$ and $\sigma_{POSt}$ that also break this same degeneracy \citep[e.g.][]{2007ApJ...657L...1S,2017MNRAS.471.4541R}. A key challenge in measuring VSPs for the SMC is that they formally require an integral over the kurtosis to infinity. This means that the kurtosis needs to be reliably measured at large radii where it could be strongly impacted by tides \citep[e.g.][]{2020MNRAS.495...98D}. For mildly disrupting dwarfs, \citet{2018MNRAS.481..860R} show that an unbiased estimate of the VSPs can still be obtained by extrapolating the kurtosis to large radii from constraints further in. Using a similar approach, we were able to recover VSP1 for both the mock and real SMC data, but VSP2 -- that is more sensitive to data further out -- was very poorly constrained and so we do not use it in this paper.

\textsc{GravSphere} has evolved significantly since its first tests in \citet{2017MNRAS.471.4541R} and \citet{2018MNRAS.481..860R}. The version we used here is the one presented in \citet{2021MNRAS.505.5686C}. The most important changes were put in place to counteract a small bias in the density beyond $\sim 4R_{1/2}$ \citep{2017MNRAS.471.4541R,2021MNRAS.501..978R}, and biases introduced by the binning method in the presence of a small number of tracers or large velocity errors \citep{2020MNRAS.496.1092G,2021A&A...651A..80Z,2021MNRAS.505.5686C}. These prompted the development of the \textsc{binulator} as a separate code to handle the data binning, and a switch of the mass model from a \textquote{non-parametric} series of power laws centred on radial bins to the \textsc{coreNFWtides} profile \citep{2018MNRAS.481..860R,2019MNRAS.487.5799R}. The two profiles have been shown to yield constraints on the cumulative mass profile that are statistically consistent with one another \citep{2020JCAP...09..004A}. However, the \textsc{coreNFWtides} density profile, $\rho_{\rm cNFWt}(r)$, has the advantage of producing profiles that more easily connect to parameters of interest in cosmological models:

\begin{equation}
\rho_{\rm cNFWt}(r) = 
\left\{
\begin{array}{ll}
\rho_{\rm cNFW} & r < r_t \\
\rho_{\rm cNFW}(r_t) \left(\frac{r}{r_t}\right)^{-\delta} & r > r_t 
\end{array}
\right.
\label{eqn:coreNFWtides}
\end{equation}
where $r_t$ sets the radius at which mass is tidally stripped from the galaxy, $\delta$ sets the logarithmic density slope beyond $r_t$; $\rho_{\rm cNFW}$ is given by:
\begin{equation} 
\rho_{\rm cNFW}(r) = f^n \rho_{\rm NFW} + \frac{n f^{n-1} (1-f^2)}{4\pi r^2 r_c} M_{\rm NFW};
\label{eqn:rhocNFW}
\end{equation}
\begin{equation}
M_{\rm cNFW}(<r) = M_{\rm NFW}(<r) f^n;
\label{eqn:coreNFW}
\end{equation}
the function $f^n$ generates a shallower profile below a core-size parameter, $r_c$: 
\begin{equation} 
f^n = \left[\tanh\left(\frac{r}{r_c}\right)\right]^n;
\end{equation}
and $M_{\rm NFW}(<r)$ is the cumulative mass of the `Navarro, Frenk \& White' (NFW) profile \citep{1996ApJ...462..563N}:
\begin{equation} 
M_{\rm NFW}(r) = M_{200} g_c \left[\ln\left(1+\frac{r}{r_s}\right) - \frac{r}{r_s}\left(1 + \frac{r}{r_s}\right)^{-1}\right];
\label{eqn:MNFW}
\end{equation}
\begin{equation}
g_c = \frac{1}{{\rm log}\left(1+c_{200}\right)-\frac{c_{200}}{1+c_{200}}};
\end{equation}
\begin{equation} 
r_{200} = \left[\frac{3}{4} M_{200} \frac{1}{\pi \Delta \rho_{\rm crit}}\right]^{1/3};
\label{eqn:r200}
\end{equation} 
where $c_{200}$ is the dimensionless {\it concentration parameter}; $\Delta = 200$ is the over-density parameter; $\rho_{\rm crit} = 136.05$\,M$_\odot$\,kpc$^{-3}$ (in a $\Lambda$CDM cosmology) is the critical density of the Universe at redshift $z=0$; $r_{200}$ is the virial radius at which the mean enclosed density is $\Delta \times \rho_{\rm crit}$; and $M_{200}$ is the virial mass -- the mass within $r_{200}$.

In \textsc{GravSphere}, the cumulative mass profile $M(r)$ is given by the sum of the stellar mass profile $M_*(r)$, which is assumed to follow the tracer distribution with a flat prior on the total stellar mass $3.45\times10^8 M_{\odot} < M_* < 5.75\times10^8 M_{\odot}$ \citep{2012AJ....144....4M}, and the \textsc{coreNFWtides} profile (equation \ref{eqn:coreNFWtides}) for the dark matter. Our priors on the model parameters are reported in the first six rows of Table~\ref{tab:parampriors}.
\begin{table}
    \centering
    \begin{tabular}{c|c|c|c|c|c}
    \hline
         & & \multicolumn{2}{c|}{Simulation run} & \multicolumn{2}{c}{Observation run} \\
        \hline
         & Parameter & Minimum & Maximum & Minimum & Maximum \\
        \hline
        1 & $\rm{log_{10}}(M_{200})$ & 5.5 & 11.5 & 7.5 & 11.5 \\
        2 & $c_{200}$ & 1.0 & 100.0 & 7.43 & 52.63 \\
        3 & $\rm{log_{10}}(r_c)$ & 0.01 & 100.0 & 0.01 & 10.0 \\
        4 & $n$ & -1.0 & 1.0 & -1.0 & 1.0 \\
        5 & $\rm{log_{10}}(r_t)$ & 0.1 & 10.0 & 1.0 & 20.0 \\
        6 & $\delta$ & 3.01 & 8.0 & 3.01 & 8.0 \\
        \hline
        7 & $\frac{M_{1,2,3}}{M_*}$ & -100 & 100 & -5 & 5 \\
        8 & $a_{1,2,3}$ & 0.01 & 2.0 & 0.1 & 2.5 \\
        \hline
        9 & $\rm{log}(r_0)$ & -2.0 & 1.0 & -2.0 & 0.0 \\
        10 & $\eta$ & 1.0 & 3.0 & 1.0 & 3.0 \\
        11 & $\beta_0$ & -0.1 & 0.1 & -0.3 & 0.3 \\
        12 & $\beta_{\infty}$ & -0.1 & 1.0 & -0.3 & 1.0 \\
        \hline
        13 & $\mu_{\nu}$ & -50 & 50 & -50 & 50 \\
        14 & $\alpha_{\nu}$ & 4.0 & 15.0 & 10 & 60.0 \\
        15 & $\beta_{\nu}$ & 1.0 & 5.0 & 1.0 & 5.0 \\
        \hline
        16 & $A_1$ & 0.0001 & 0.35 & 0.0001 & 0.35 \\
        17 & $\mu_1$ & 20.0 & 95.0 & 20.0 & 95.0 \\
        18 & $\sigma_1$ & 30.0 & 125.0 & 30.0 & 125.0 \\
        19 & $A_2$ & 0.0001 & 0.35 & 0.0001 & 0.35 \\
        20 & $\mu_2$ & -95.0 & -20.0 & -95.0 & -20.0 \\
        21 & $\sigma_2$ & 30.0 & 125.0 & 30.0 & 125.0 \\
        \hline
    \end{tabular}
    \caption{Bounds on the priors and parameters used by \textsc{GravSphere} and the \textsc{binulator}.
    \textit{The first six rows} are the bounds of the flat priors assumed for the mass profile. $M_{200}$ is in $M_{\odot}$, $c_{200}$, $n$ and $\delta$ are dimensionless, $r_c$ and $r_t$ are in kpc.
    \textit{Rows 7 and 8} are the priors of the Plummer spheres used to fit the tracers density, with the $M_j/M_*$ dimensionless (as the surface density data are normalised to integrate to 1) and $a_j$ in kpc.
    \textit{Rows from 9 to 12} are the priors for the anisotropy profile parameters where $r_0$ is in kpc and $\eta$, $\beta_0$, and $\beta_{\infty}$ are dimensionless.
    \textit{Rows 13 to 15} are the bounds of the parameter space searched for the fit of the residual velocity distributions.
    \textit{Rows 16 to 21} are the parameters of the two secondary Gaussians used to model the contaminants of the velocity distributions (a standard Gaussian of amplitude $A_i$, mean $\mu_i$, and dispersion $\sigma_i$).
    }.
    \label{tab:parampriors}
\end{table}

The tracer density is given by a sum of a series of $N_p$ Plummer spheres \citep{1911MNRAS..71..460P,2016MNRAS.459.3349R}:
\begin{equation}\label{tracers_sum}
    \nu=\sum_j^{N_p}\frac{3M_j}{4\pi a_j^3}{\bigg(1+\frac{r^2}{a_j^2}\bigg)}^{-5/2}
\end{equation}
where $M_j$ and $a_j$ are the mass and scale length of each individual component. $\Sigma_*(R)$ appears in Eqs.~\ref{jeans_true}, ~\ref{posradial}, and ~\ref{postang} and has to be compared with the data. Eq.~\ref{tracers_sum} makes $\Sigma_*(R)$ analytic:
\begin{equation}\label{surfprof1}
    \Sigma_*(R)=M_*\sum_j^{N_p}\frac{M_j}{\pi a_j^2}{\big(1+\frac{R^2}{a_j^2}\big)}^{-2}
\end{equation}
Typically, $N_p=3$ is enough to model the tracer density, especially since the masses are allowed to be negative (under a constraint that the total density at all radii remains positive; \citealt{2016MNRAS.459.3349R}). Rows 7 and 8 of Table~\ref{tab:parampriors} report the parameter space that the code searched when fitting the tracer densities for the models. The \textsc{binulator} does a first fit of the tracer density and then \textsc{GravSphere} searches for a new solution around the best fit within a preset tolerance (here chosen to be $10^{-3}$, as the data are very constraining).

The velocity anisotropy profile follows \citet{2007A&A...471..419B} and \citet{2017MNRAS.471.4541R}:
\begin{equation}\label{normal_anisotropy}
    \beta(r)=\beta_0+(\beta_{\infty}-\beta_0)\frac{1}{1+{(\frac{r_0}{r})}^{\eta}}
\end{equation}
where $\beta_0$ is the anisotropy at the centre, $\beta_{\infty}$ is the value at infinity, $r_0$ is the transition radius, $\eta$ dictates the steepness of the profile, and the priors for all the parameters are given in rows 9 to 12 of Table~\ref{tab:parampriors}. This definition of anisotropy allows for a wide range of anisotropy profiles while making Eq.~\ref{gierre} analytic. Even more general analytic forms are discussed and presented in \citealt{2017MNRAS.471.4541R}.

$\beta(r)$, as defined above, has values over an infinite range ($-\infty < \beta < 1$) which is problematic for model fitting and, hence, \textsc{GravSphere} uses instead a symmetrised version \citep{2006MNRAS.367..387R}:
\begin{equation}
    \widetilde{\beta}=\frac{\sigma_r^2-\sigma_t^2}{\sigma_r^2+\sigma_t^2}=\frac{\beta}{2-\beta}
\end{equation}
With this definition, the anisotropy is bounded on a finite parameter space: $\widetilde{\beta}=1$ is full radial anisotropy, $\widetilde{\beta}=0$ is isotropy and $\widetilde{\beta}=-1$ is full tangential anisotropy.

\textsc{GravSphere} fits the tracers' surface density (Eq.~\ref{surfprof1}) and velocity dispersion profiles  (Eq.~\ref{jeans_true}, Eqs.~\ref{posradial} and ~\ref{postang}) using the MCMC code \textsc{Emcee} \citep{2013PASP..125..306F}. These fits allow for the reconstruction of the dark matter density and stellar velocity anisotropy profiles through the other equations presented here.

\begin{figure*}
\centering
\includegraphics[width=\textwidth]{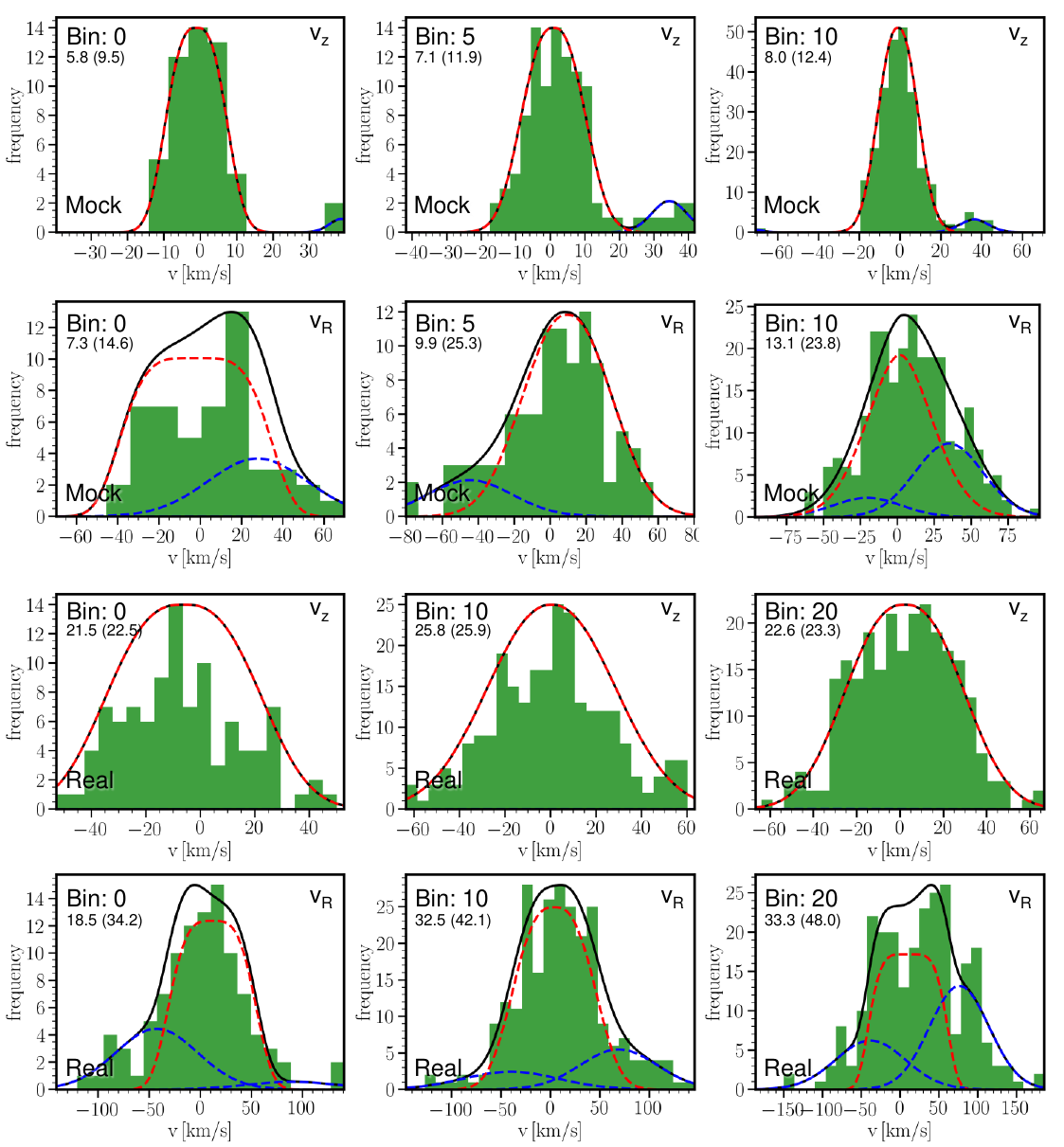}
\caption{The velocity distribution functions (VDFs) of some of the bins of the heavily disrupted SMC-analogue (top two rows) and of the real SMC (bottom two rows). In each panel, the VDFs are the green histograms, the best \textsc{binulator} fit to the distribution (including both the main SMC generalised Gaussian and the secondaries representing the debris) the solid black line, the fit to the SMC the dashed red line and the secondary distributions for the debris the dashed blue lines. The numbers reported below the bin identifier in the top left corner of each panel are the dispersion of the SMC distribution: the first number is the dispersion recovered by the \textsc{binulator} fit, the number in brackets is the dispersion that would be recovered by fitting the entire bin with a single component (so without any distinction between SMC and debris). \textit{First and third row:} In the VDFs for the LOS velocity distribution, $V_z$, the debris are clearly identifiable and removed for the simulation (first row) and do not affect the real SMC data distribution (third row). \textit{Second and fourth row:} the VDFs along the radial direction on the plane of the sky, $V_R$, the debris are more difficult to disentangle from the bulk SMC distribution for both simulation (second row) and real SMC data (fourth row). The tangential direction on the plane of the sky, $V_T$, is omitted as it shows similar behaviour to $V_R$. \label{fig:example_binulator}}
\end{figure*}

\subsection{The \textsc{binulator}}\label{binu}

In \citet{2021MNRAS.505.5686C}, the binning routines of \textsc{GravSphere} were redeveloped and built into their own code, the \textsc{binulator}. This implements one main change: each projected radial bin (that contains an equal number of stars, weighted by their membership probability) is fit with a generalised Gaussian, convolved with the error PDF of each star:
\begin{equation}\label{binulatorpdf}
    p_i=\frac{\beta_{\nu}}{2\tilde{\alpha}_{\nu}\Gamma(1.0/\beta_{\nu})}\rm{exp}\bigg(-{\frac{|\nu_i-\mu_{\nu}|}{\tilde{\alpha}_{\nu}}}^{\beta_{\nu}}\bigg)
\end{equation}
where:
\begin{equation}\label{convolved}
    \widetilde{\alpha}^2=\alpha_{\nu}^2+\sigma_{e,i}^2\frac{\Gamma(1.0/\beta_{\nu})}{\Gamma(3.0/\beta_{\nu})}
\end{equation}
and $\sigma_{e,i}$ is the width of the PDF of the error on the $i$-th star; $\nu_i$ is the velocity (be it line-of-sight or along one of the plane of the sky directions) of the $i$-th star; $\Gamma$(x) is the Gamma Function; and $\mu_{\nu}$, $\alpha_{\nu}$ and $\beta_{\nu}$ are parameters fit to each bin. These allow us to recover the mean, $\mu_{\nu}$, variance, $\sigma_{\nu}^2=\alpha_{\nu}^2\Gamma(3.0/\beta_{\nu})/\Gamma(1.0/\beta_{\nu})$ and kurtosis, $k=\Gamma(5.0/\beta_{\nu})\Gamma(1.0/\beta_{\nu})/[\Gamma(3.0/\beta_{\nu})]^2$, of that bin (c.f. the similar method in \citealt{2020MNRAS.499.5806S}). Note that the above is an analytic approximation to the true convolution integral. \citet[][their Figure 10]{2021MNRAS.505.5686C} show that this matches the true convolution integral at typically better than 5\% accuracy, and rarely more poorly than 10\%. Rows 13 to 15 of Table~\ref{tab:parampriors} list the priors used for the fit of the velocity PDFs for the real and simulated data.

\subsection{Removing tidal debris}\label{sec:binulatormock}

Due to the heavy tidal disruption, the SMC is currently undergoing, the kinematic sample is likely contaminated by unbound debris which invalidates the assumption of dynamical equilibrium required for the modelling. This is a problem long-recognised in the literature \citep[i.e.][]{2007MNRAS.378..353K}, made particularly challenging by the fact that the debris stars are chemically similar to the bound stars. A standard solution is to sigma clip stars with anomalously high velocities \citep[i.e.][]{2004ApJ...611L..21W,2007MNRAS.378..353K}. However, this raises the spectre that genuine member stars are accidentally removed, impacting estimates of both the dispersion and, in particular, the kurtosis that is sensitive to the wings of the velocity distribution function. Furthermore, it is hard to marginalise over ambiguous stars that may or may not be bound, since they must be either in or out.

Here, we use the {\sc binulator} to remove the tidal debris by representing the debris with up to two secondary Gaussians (bounds on their parameters in rows 16 to 21 of Table~\ref{tab:parampriors}) that we add to the velocity distribution function in equation \ref{binulatorpdf}. These secondary Gaussians are forced to have a mean distinct from zero (one positive and one negative, for debris on each side of the main velocity distribution), and can have a negligible amplitude. These choices ensure the debris distributions are separate from the bulk motion of the SMC (at mean zero), and also that the secondary Gaussians can be present or absent independently from each other in each bin. This allows us to fully marginalise over the bound member stars when {\sc binulator} fits its velocity distribution function bin by bin. The {\sc binulator} transforms the observed velocities on the plane of the sky in the radial and tangential components with respect to the centre of the SMC, so ($V_r$,$\mu_{\alpha}$,$\mu_{\delta}$) becomes ($V_z$,$V_R$,$V_T$), and the velocity distribution in each direction is fitted independently. We test this idea using mock data drawn from our heavily disrupted simulated SMC-analogue in \S\ref{sec:binulatortest}. An example fit of {\sc binulator}'s \textquote{generalised Gaussian + secondary Gaussians} PDF to some of the bins for the heavily disrupted mock (top two rows) and the real SMC (bottom two rows) is shown in Fig. \ref{fig:example_binulator}. In each panel, the velocity distribution is in green, the sum of the Gaussians is the solid black line, the fit to the SMC distribution is the dashed red line and the secondary Gaussians for the debris are the blue dashed lines. The numbers in the top left corner of each panel, below the bin identifier, are the dispersion for that bin recovered by the \textsc{binulator} and, in brackets, the dispersion recovered by fitting a single component to the whole bin (thus without any distinction between the distributions of the main SMC and of the debris). The difference between these two recovered values gives a gauge of the importance of the removal of the contamination due to the tidal debris. As can be seen from the figure, the debris are easily recognisable and removed in the LOS velocity distributions ($V_z$, first and third row of the figure) but are more difficult to disentangle from the bulk SMC motion in the radial component on the plane of the sky ($V_R$, second and fourth row, the tangential component $V_T$ shows the same behaviour as $V_R$). This difference in the overall effect and severity of the debris' contamination leads our model to be more tightly constrained by the LOS data and to larger errors in the estimated distributions in the plane of the sky directions. It is important to note that we do not actually use the best-fit velocity distribution function found by {\sc binulator} to generate the velocity dispersion profiles and their uncertainties, but rather we use the median and 68\% confidence intervals of the distribution of the many fits that the \textsc{binulator} does for each bin (to marginalise over model degeneracies).

\section{Results}\label{res}

\subsection{Tests on the simulated SMC-analogues}\label{tests}

Before applying our methodology (\S\ref{modeling}) to the real SMC data, we first test it on simulated mock SMC analogues. We consider two mocks, as described in \S\ref{sims}: a bound SMC that has undergone very little tidal disruption, and a heavily disrupted SMC that is close to dissolution. This latter is closer to the real situation, but likely even more extreme as the mock SMC has a starting mass lower than the original mass of the SMC (as reconstructed via abundance matching, i.e. in \citealt{2019MNRAS.487.5799R}). As such, it represents a conservative test-case. We first assess how well {\sc binulator} can remove unbound tidal debris along the line of sight (\S\ref{sec:binulatortest}); we then apply {\sc GravSphere} to the mock data to determine how well we can recover the inner dark matter density profile and stellar velocity anisotropy (\S\ref{sec:mocktests}). Other profiles recovered by our models are reported in Appendix~\ref{app_profiles} for completeness.

\subsubsection{Testing the removal of tidal debris}\label{sec:binulatortest}
\begin{figure}
\centering
\includegraphics[width=\columnwidth]{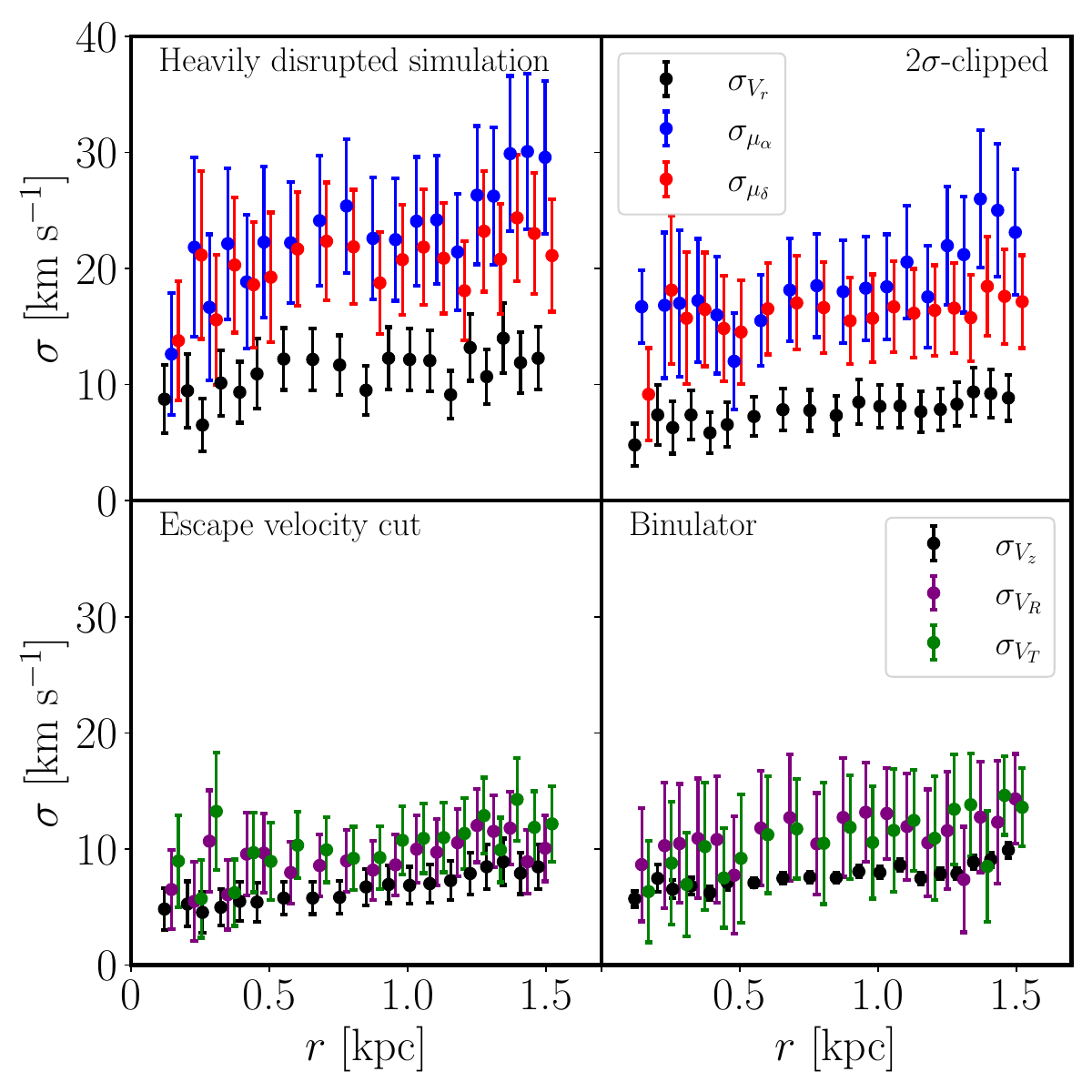}
\caption{Velocity dispersion profiles for the inner regions of the heavily disrupted simulation. The black points show the velocity dispersion along the line-of-sight ($\sigma_{V_r}$ in the top panels, $\sigma_{V_z}$ in the bottom panels), the blue points $\sigma_{\mu\alpha}$, the red points $\sigma_{\mu\delta}$, the purple points $\sigma_{V_R}$, and the dark green points $\sigma_{V_T}$. \textit{Top left panel}: velocity dispersions for the full data sample. \textit{Top right panel}: Velocity dispersions after $2\sigma$-clipping the velocity distributions, which are assumed to be Gaussians (see Fig.~\ref{fig:kinematics_distributions}). \textit{Bottom left panel}: velocity dispersions after iterated escape velocity cuts. \textit{Bottom right panel}: velocity dispersions after the decontamination done by the \textsc{binulator}. Inside each panel the bins are the same but are displaced artificially along the X-axis for clarity. \label{fig:full_2s_escvel_comparison}}
\end{figure}

In Fig.~\ref{fig:full_2s_escvel_comparison}, we show the velocity dispersion profiles derived from the heavily disrupted simulation. In the top left panel, we show the results including all data, both bound and unbound stars. In the top right panel, we show results for the same simulation but clipping all data beyond 2 standard deviations from the dispersion derived for each bin (2$\sigma$-clipped), assuming the original velocity distributions to be Gaussians (see Fig.~\ref{fig:kinematics_distributions}). In the bottom left panel, we show the results obtained by removing the stars with a velocity above the escape velocity $V_{esc}$ at their respective position (i.e. the unbound stars).
\begin{equation}\label{escapevelocity}
    V_{esc}=\sqrt{-2 \Phi(r)}
\end{equation}
\begin{equation}\label{precisePhi}
    \Phi(r)=-G\int_{r}^{\infty}\frac{M(<r)}{r'^2}dr'
\end{equation}
with $\Phi(r)$ the gravitational potential, $M(r)$ tabulated from the particle data to a very high distance and the integral estimated numerically.
Due to the advanced stage of tidal disruption of the simulation and the discrete mass distribution of the simulation particles, we iterated the selection process until it stopped removing particles.
Finally, in the bottom right panel of Fig.~\ref{fig:full_2s_escvel_comparison}, we show the profiles derived by the \textsc{binulator}. The colours of the points in the bottom panels are different because the \textsc{binulator} transformed the proper motions into the radial and tangential velocities on the plane of the sky and we did the same for the escape velocity cut sample, for ease of comparison (the better kinematically behaved, i.e. bound, sample will always have lower values of the velocity dispersions anyway). As can be seen from the bottom left panel, when removing the stars with a velocity higher than the escape velocity, the inner dispersion profiles become consistent with one another: the inner velocity dispersion of bound stars is isotropic. Sigma clipping of the data in each bin (top right panel) is unable to reproduce this behaviour, with the 2$\sigma$-clipped dispersions remaining significantly tangentially anisotropic, even in the innermost bin. The \textsc{binulator} (bottom right panel), however, is able to recover the correct behaviour within its statistical uncertainties by removing the unbound stars. The slight rise outward in the dispersions, along all velocities, that both the escape velocity cut sample (bottom left panel) and the \textsc{binulator} (bottom right panel) exhibit is likely due to the physical onset of tides \citep{2006MNRAS.366..429R,2006MNRAS.367..387R}.

\subsubsection{Testing the mass modelling}\label{sec:mocktests}
\begin{figure*}
     \centering
     \begin{subfigure}[b]{0.45\textwidth}
         \centering
         \includegraphics[width=\textwidth, height=70mm]{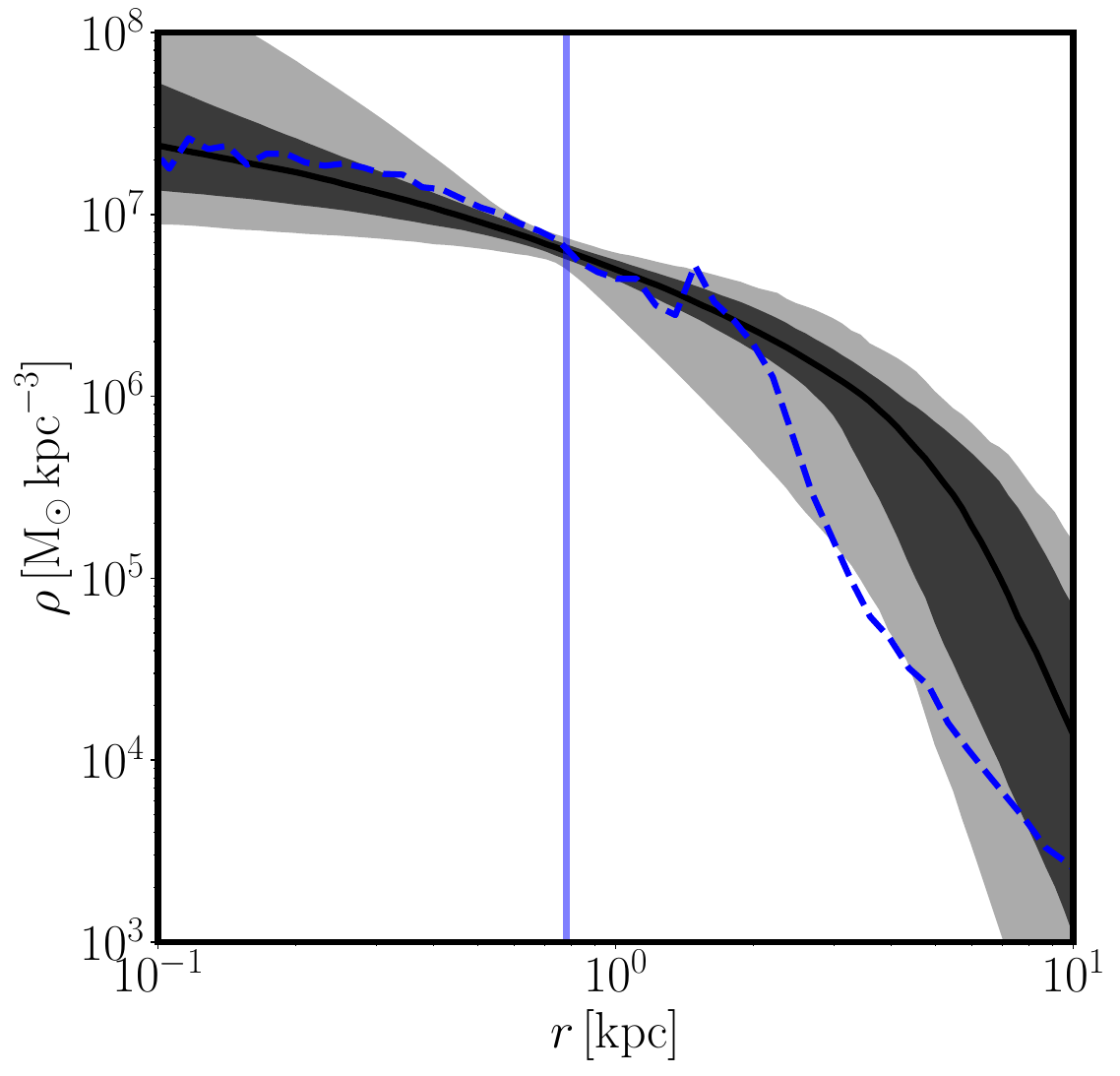}
     \end{subfigure}
     \begin{subfigure}[b]{0.45\textwidth}
         \centering
         \includegraphics[width=\textwidth, height=70mm]{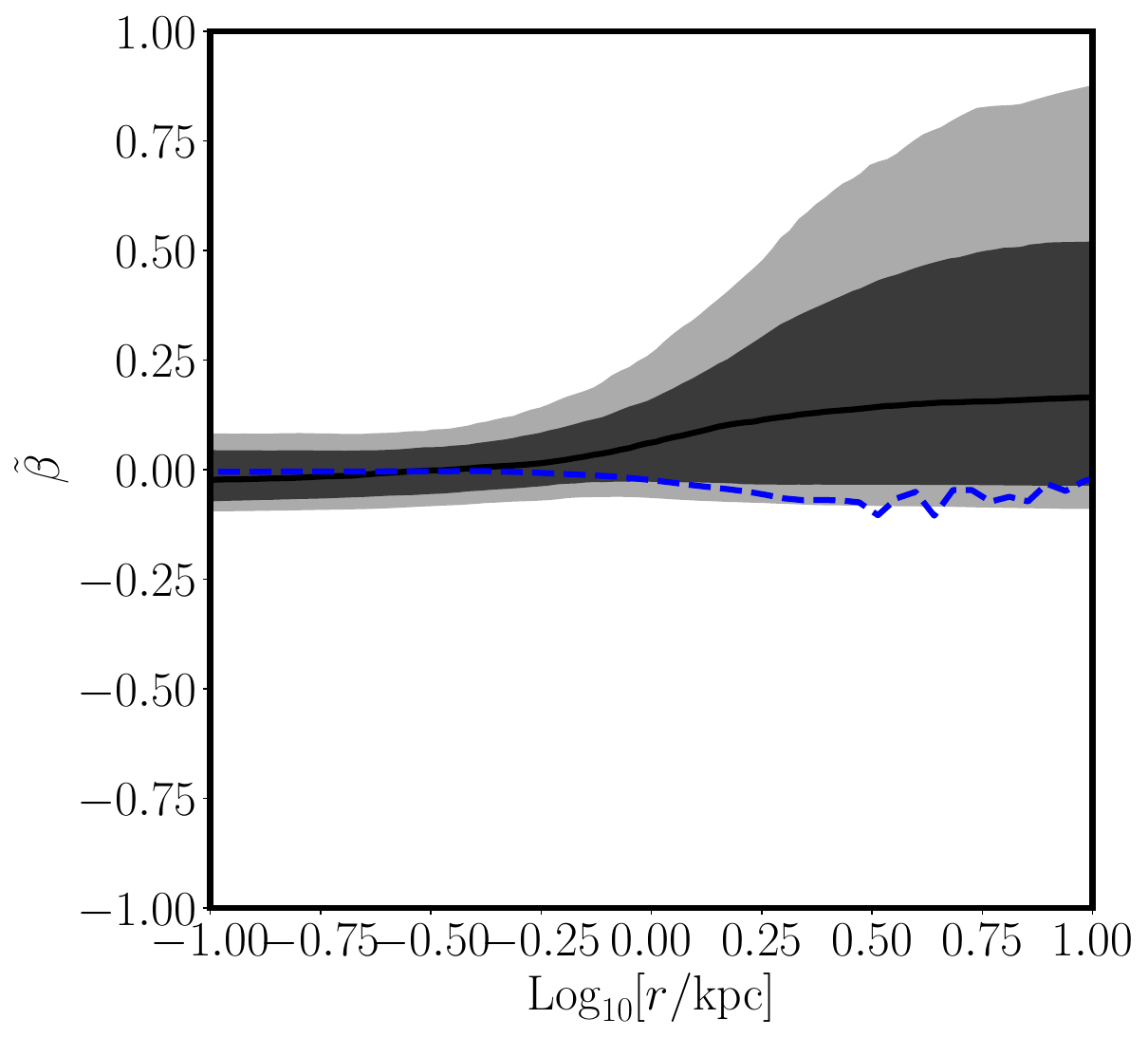}
     \end{subfigure}\\
	\centering
     \begin{subfigure}[b]{0.45\textwidth}
         \centering
         \includegraphics[width=\textwidth, height=70mm]{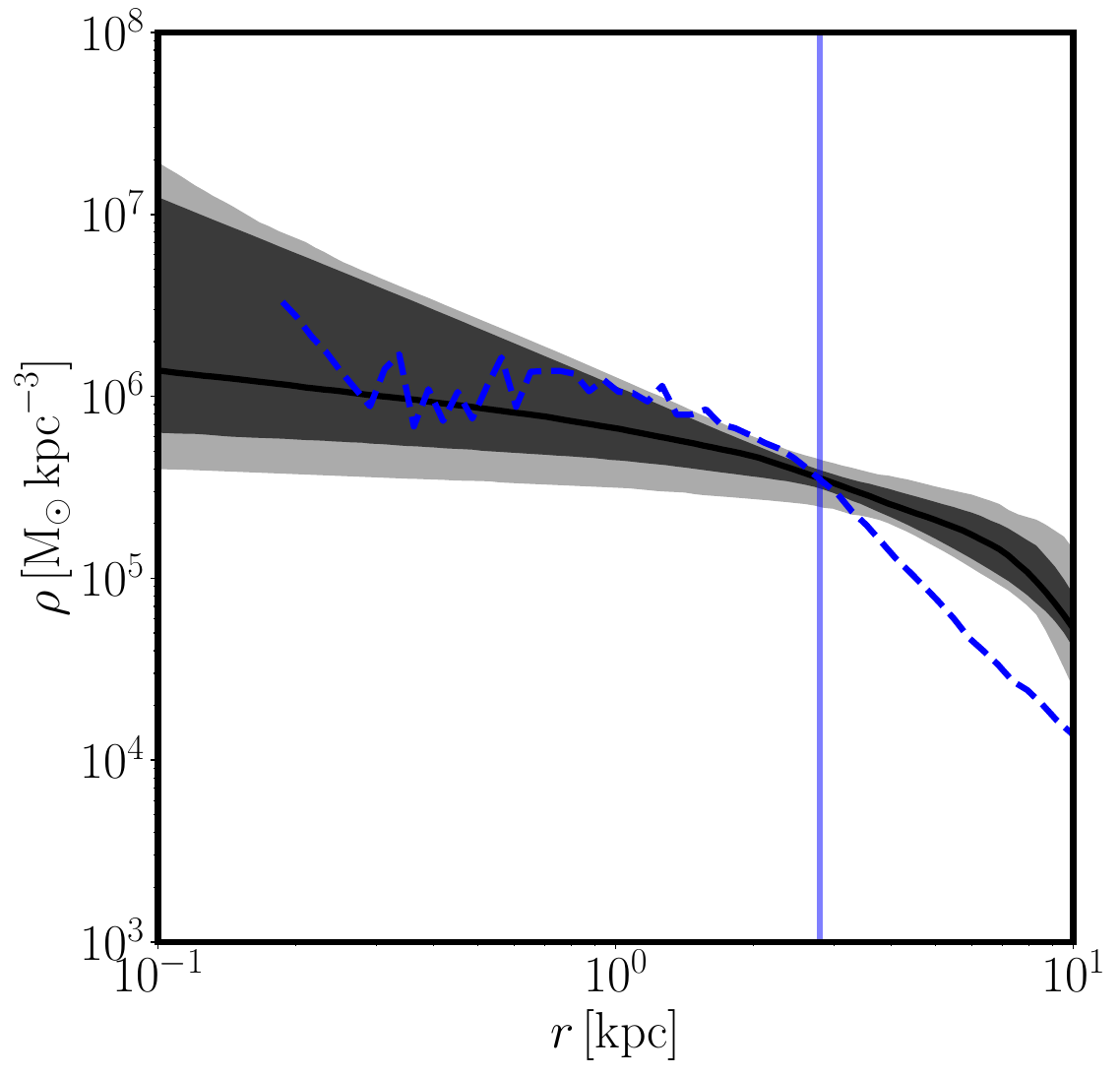}
     \end{subfigure}
     \begin{subfigure}[b]{0.45\textwidth}
         \centering
         \includegraphics[width=\textwidth, height=70mm]{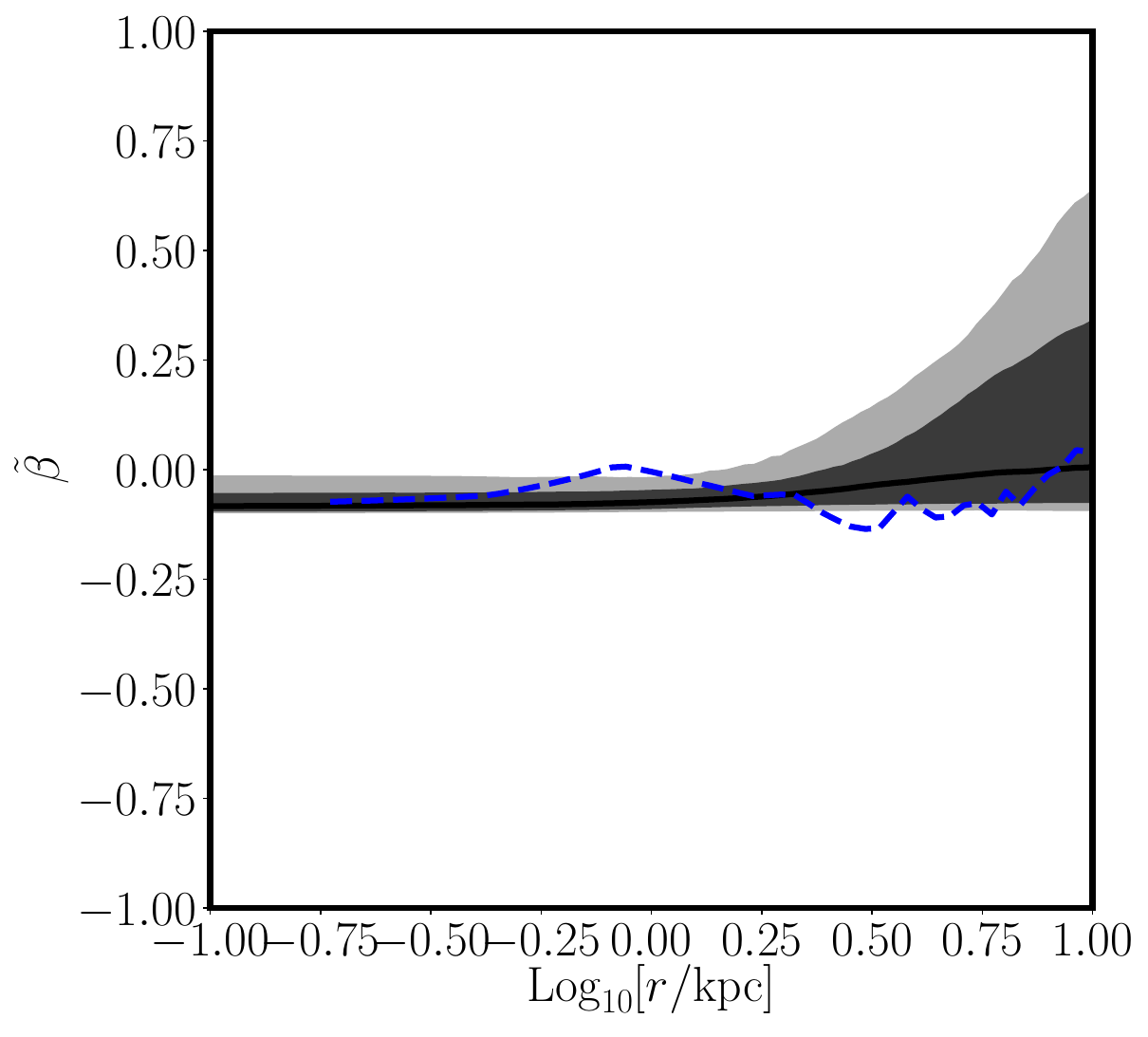}
     \end{subfigure}
        \caption{\textit{Left panels}: recovered mass density profile, $\rho(r)$, of the bound (top) and heavily disrupted (bottom) simulations. The black lines and grey contours mark the median, 68\% (dark grey), and 95\% (light grey) confidence intervals of the \textsc{GravSphere} fit, respectively. The blue dashed lines show the profiles computed from the simulation data. The blue vertical lines mark the half-stellar mass radius computed by \textsc{binulator}. \textit{Right panels}: recovered symmetrised anisotropy profile $\widetilde{\beta}(r)$ for the bound (top) and heavily disrupted (bottom) simulations. The solid black lines, grey contours, vertical blue lines, and blue dashed lines are as in the left panels.}
        \label{fig:multi_bound}
\end{figure*}

In this section, we now apply \textsc{GravSphere} to model the mock data surface brightness and velocity dispersion profiles extracted using \textsc{binulator} (\S\ref{sec:binulatormock}). The results for the bound simulation that has not experienced any significant tidal forces are shown in the top row of Fig.~\ref{fig:multi_bound} while the bottom row is for the heavily disrupted simulation. The left and right columns of the figure show the recovered median (black line), 68\% (dark  grey), and 95\% (light grey) confidence intervals for $\rho(r)$ and $\widetilde{\beta}(r)$, respectively, as compared to the true solutions (blue data points and dashed lines). As can be seen, {\sc GravSphere} correctly recovers all three within its 95\% confidence intervals.

Fig.~\ref{fig:multi_bound} shows that {\sc GravSphere} is able to recover the density and velocity anisotropy profiles within its 95\% confidence intervals for both the bound simulation (top panels) and the heavily disrupted simulation (bottom panels) within the half-stellar mass radius (vertical blue line, $R_{1/2}$). Beyond $R_{1/2}$, the recovered density profile is biased high for the heavily disrupted simulation as compared to the true solution (see Fig.~\ref{fig:multi_bound}, bottom left panel), while the velocity anisotropy profile also fluctuates slightly outside of the 95\% confidence intervals (bottom right panel). This behaviour is to be expected given that the heavily disrupted simulation becomes unbound beyond $R_{1/2}$, with the percentage of bound stars quickly dropping below $90\%$ outside 1.7\,kpc.

\subsection{Mass modelling of the real SMC}\label{real_smc}

In this section, we show and discuss the principal results of our modelling of the SMC, namely the successful decontamination with the \textsc{binulator}, the recovery of the mass density and velocity anisotropy profile and further insights derived from these two variables. Other profiles recovered by our models are reported in Appendix~\ref{app_profiles} for completeness.

\subsubsection{Removing tidal debris}

\begin{figure*}
     \centering
     \begin{subfigure}[b]{0.48\textwidth}
         \centering
         \includegraphics[width=\textwidth]{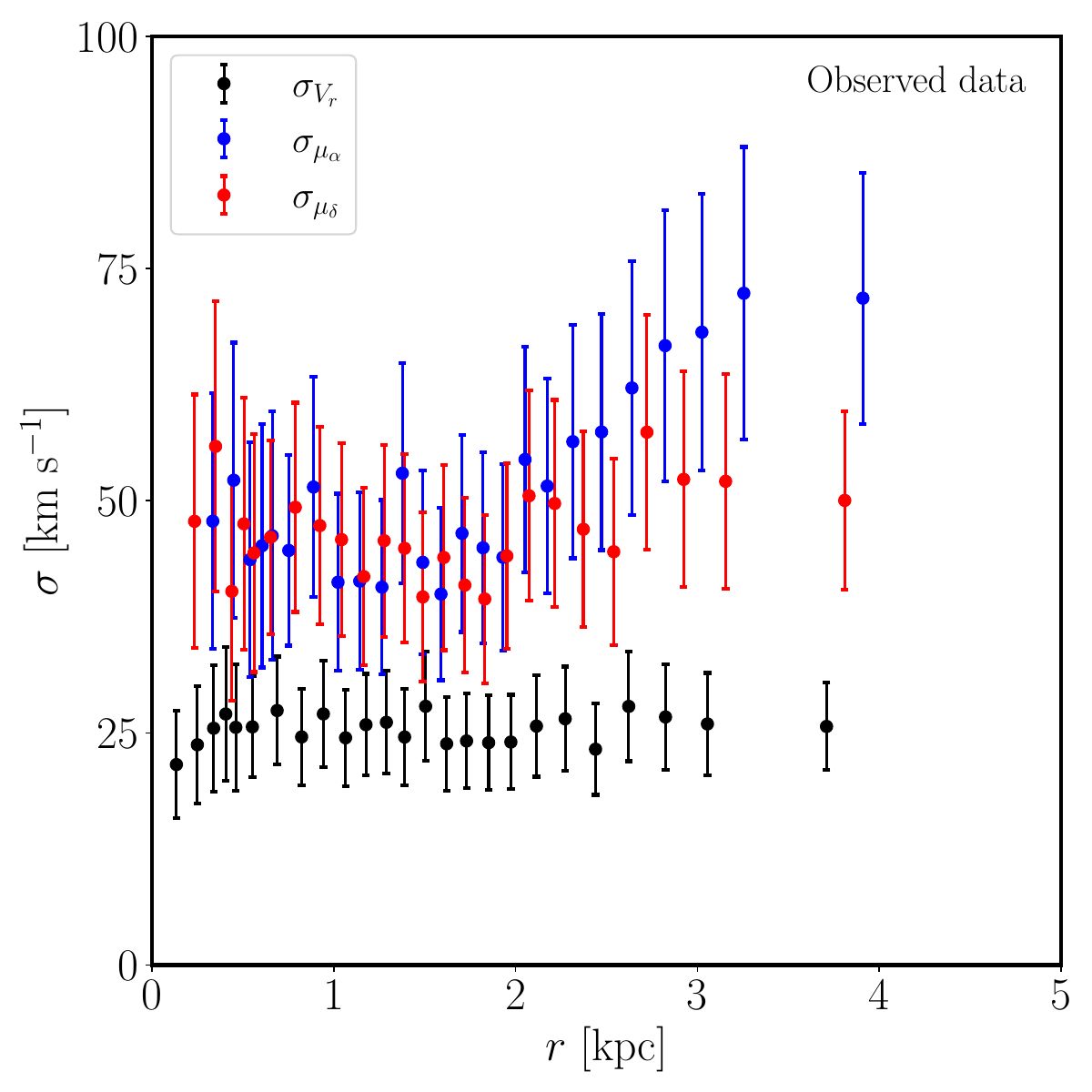}
     \end{subfigure}
     \begin{subfigure}[b]{0.48\textwidth}
         \centering
         \includegraphics[width=\textwidth]{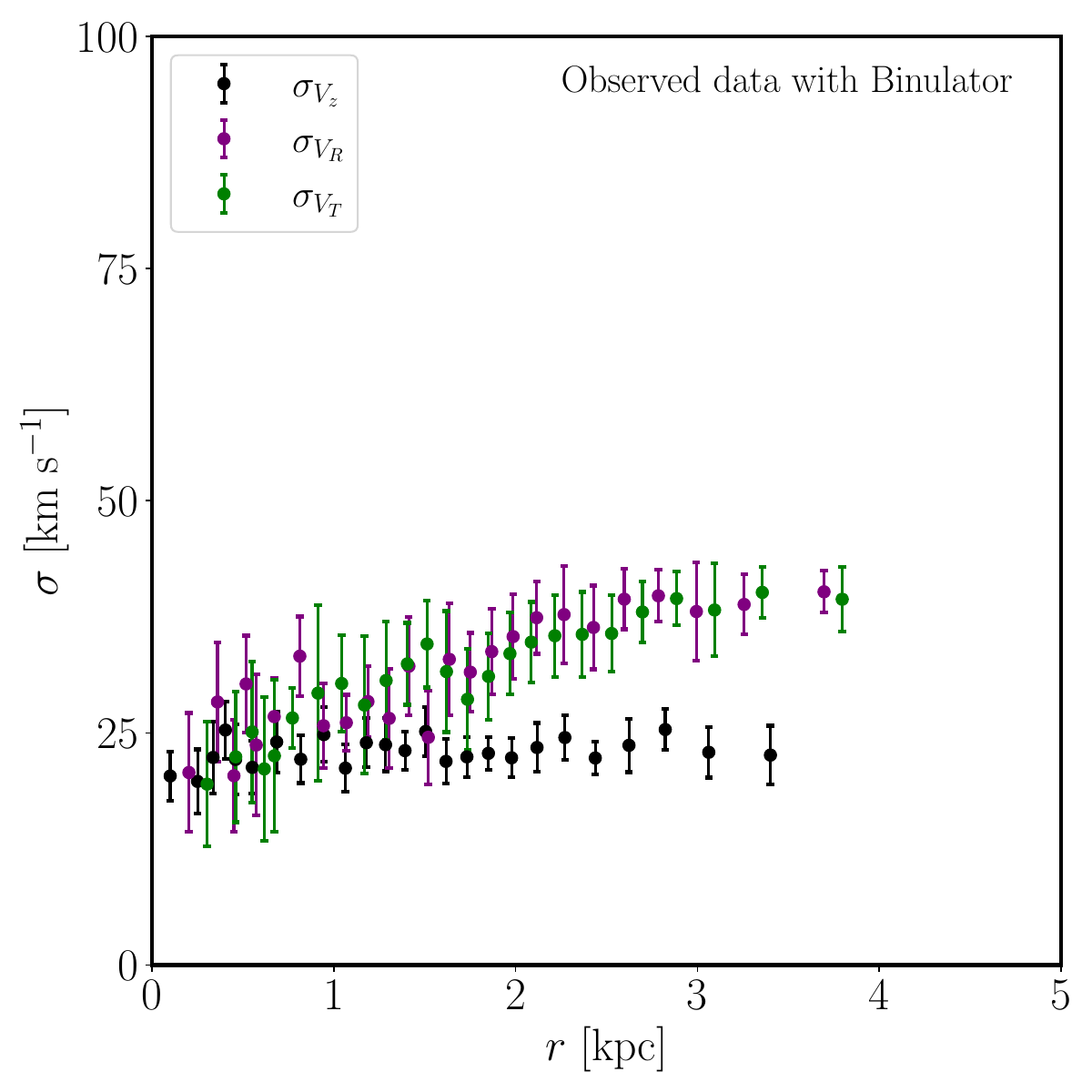}
     \end{subfigure}
        \caption{\textit{Left panel}: dispersion profiles of $V_r$ (black points), $\mu_{\alpha}$ (blue points), and $\mu_{\delta}$ (red points) for the observed data. The data are binned at the same positions but displaced along the X-axis for clarity. \textit{Right panel}: dispersion profiles of $V_z$ (black points), $V_R$ (purple points), and $V_T$ (dark green points) for the observed data after the \textsc{binulator} fit of the velocity distributions. The data are binned at the same positions but displaced along the X-axis for clarity. The rising anisotropy after $\sim1.5\,$ kpc likely owes to the physical onset of tides, rather than artificial inflation of the dispersions due to line-of-sight contamination. As such, \textsc{binulator} is unable to correct for this. For this reason, we limit our fits to within the half-mass radius.}
        \label{fig:binulator_power}
\end{figure*}

We first check the impact of the removal of tidal debris by the \textsc{binulator}. Fig.~\ref{fig:binulator_power} compares the dispersions of the data processed by the \textsc{binulator} (right) with the dispersion profiles of the observed data, taken as simple variances of each data bin (left). The decontamination has dampened the tangential anisotropy, with only some mild residual contamination likely due to the physical onset of tides \citep{2006MNRAS.367..387R} remaining after $r=1.5\,$ kpc, this is reminiscent of the behaviour of the heavily disrupted mock in Fig.~\ref{fig:full_2s_escvel_comparison} and it is the reason why we limited the \textsc{GravSphere} fitting to within the half-mass radius.

\subsubsection{The {\sc GravSphere} model of the SMC}
\begin{figure*}
     \centering
     \begin{subfigure}[b]{0.48\textwidth}
         \centering
         \includegraphics[width=\textwidth, height=75mm]{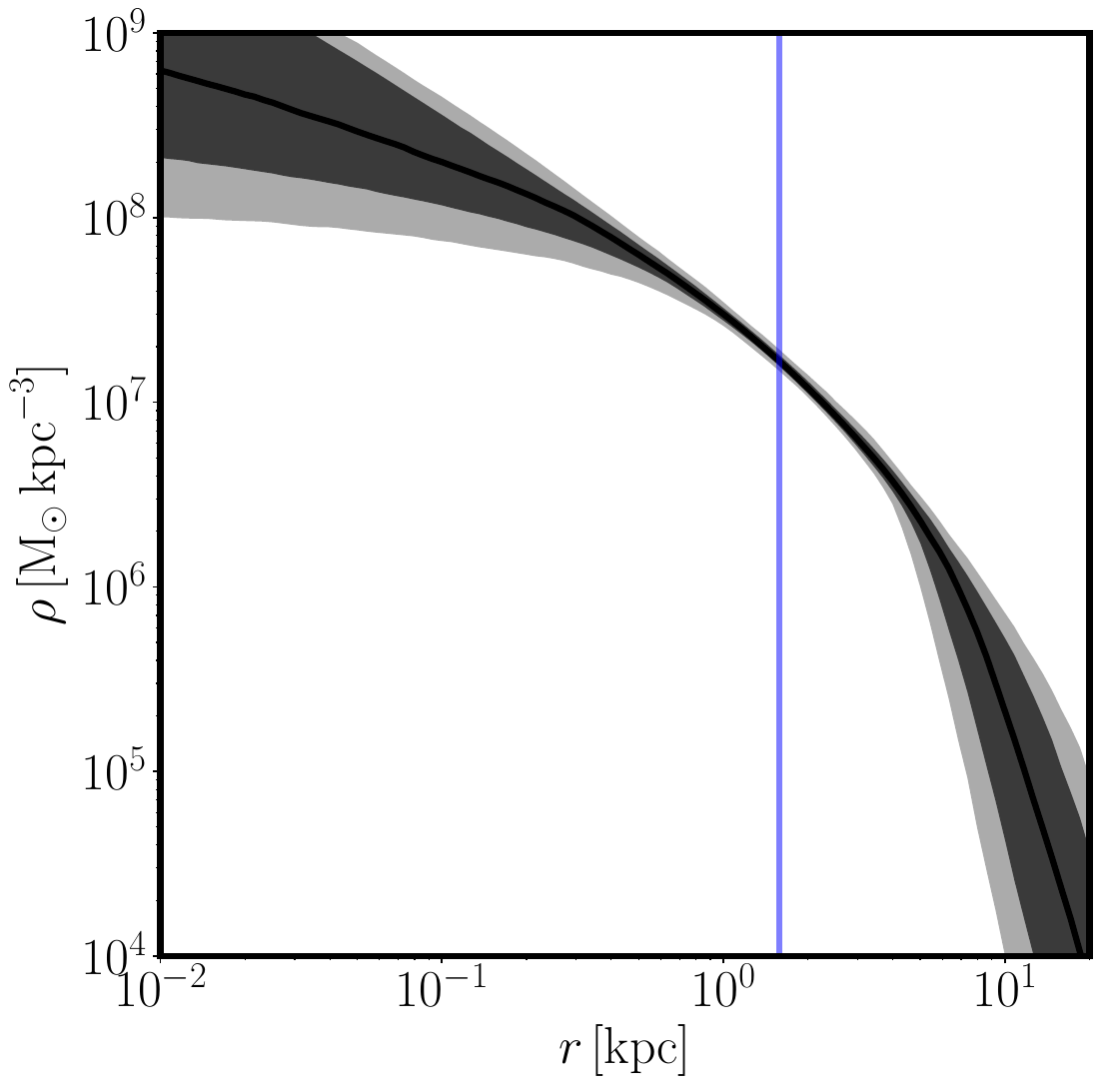}
     \end{subfigure}
     \begin{subfigure}[b]{0.48\textwidth}
         \centering
         \includegraphics[width=\textwidth, height=75mm]{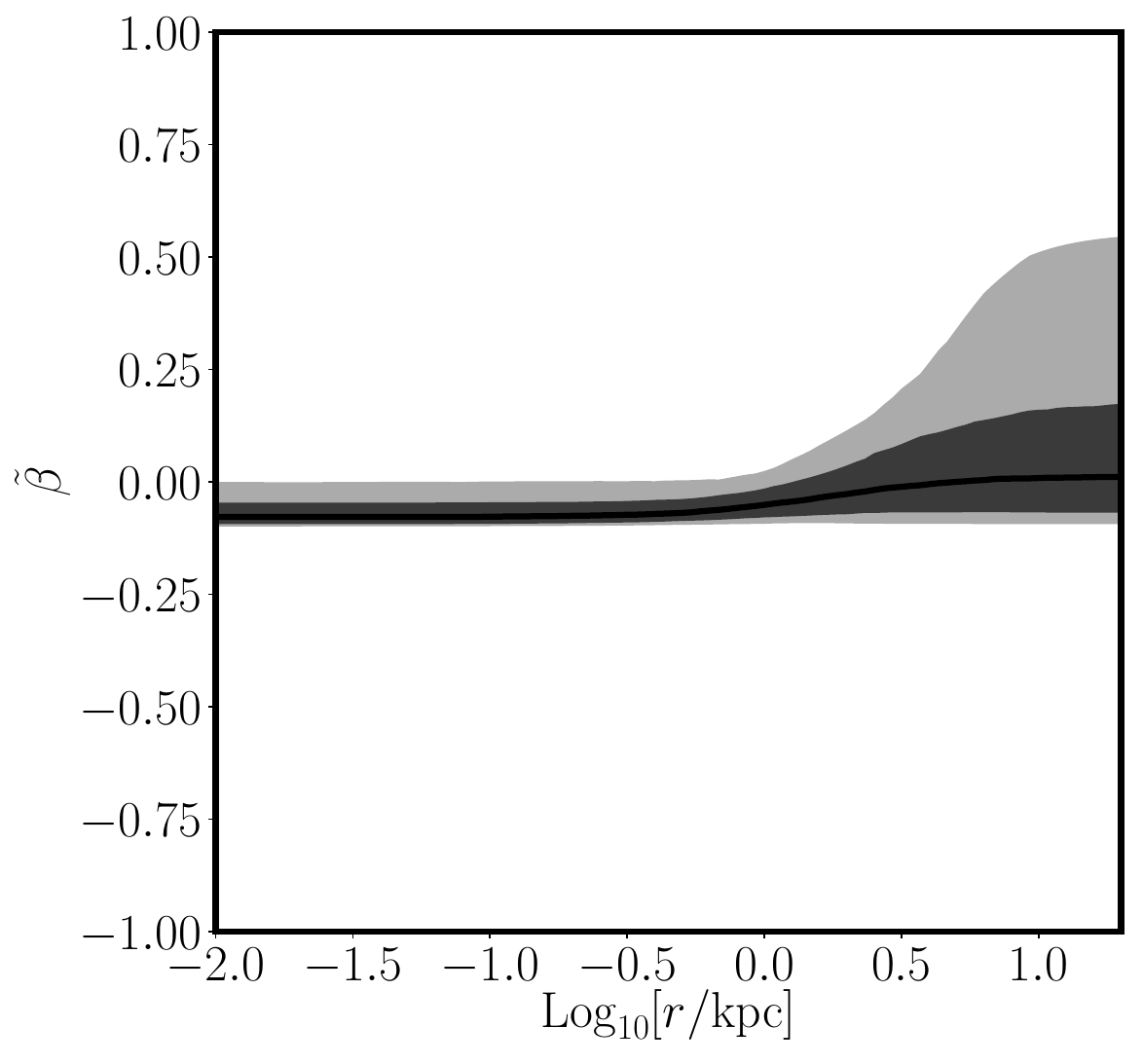}
     \end{subfigure}
        \caption{\textit{Left panel}: $\rho(r)$ profile recovered by \textsc{GravSphere} for the SMC data. The black line is the best-fit solution and the dark and light gray regions are, respectively, the 68\% and 95\% confidence intervals. The faint blue vertical line is the recovered half-light radius. \textit{Right panel}: $\widetilde{\beta}(r)$ profile recovered by \textsc{GravSphere} for the SMC data. The black line is the best-fit solution and the dark and light gray regions are, respectively, the 68\% and 95\% confidence intervals.}
       \label{fig:smc_grav}
\end{figure*}

In Fig.~\ref{fig:smc_grav}, we show the {\sc GravSphere} recovery of the dark matter density (left) and velocity anisotropy (right) profiles for the real SMC. As reflected in the data (Fig.~\ref{fig:binulator_power}), {\sc GravSphere} favours some mild tangential anisotropy, though at 95\% confidence it is consistent with being isotropic at all radii probed. The density profile is well-constrained over the range $0.5 \lesssim R/R_{1/2} \lesssim 2$ and appears more cusp-like than cored (constant density). We discuss this further in \S\ref{sec:densityprofile}.

\subsubsection{The present and pre-infall mass of the SMC}\label{sec:SMCmasS}

Regarding the mass of the SMC, the recovered density profile suggests a dark matter mass within $3$\,kpc of $M_{DM}(\leq 3\,{\rm kpc})=1.39\pm0.08 \times 10^9 M_{\odot}$ and a stellar mass within the same radius of $M_{*}(\leq 3\,{\rm kpc})=0.34\pm0.05 \times 10^9 M_{\odot}$. We compare this with other literature estimates in \S\ref{sec:discussionmass}.

\begin{figure}
\centering
\includegraphics[width=\columnwidth]{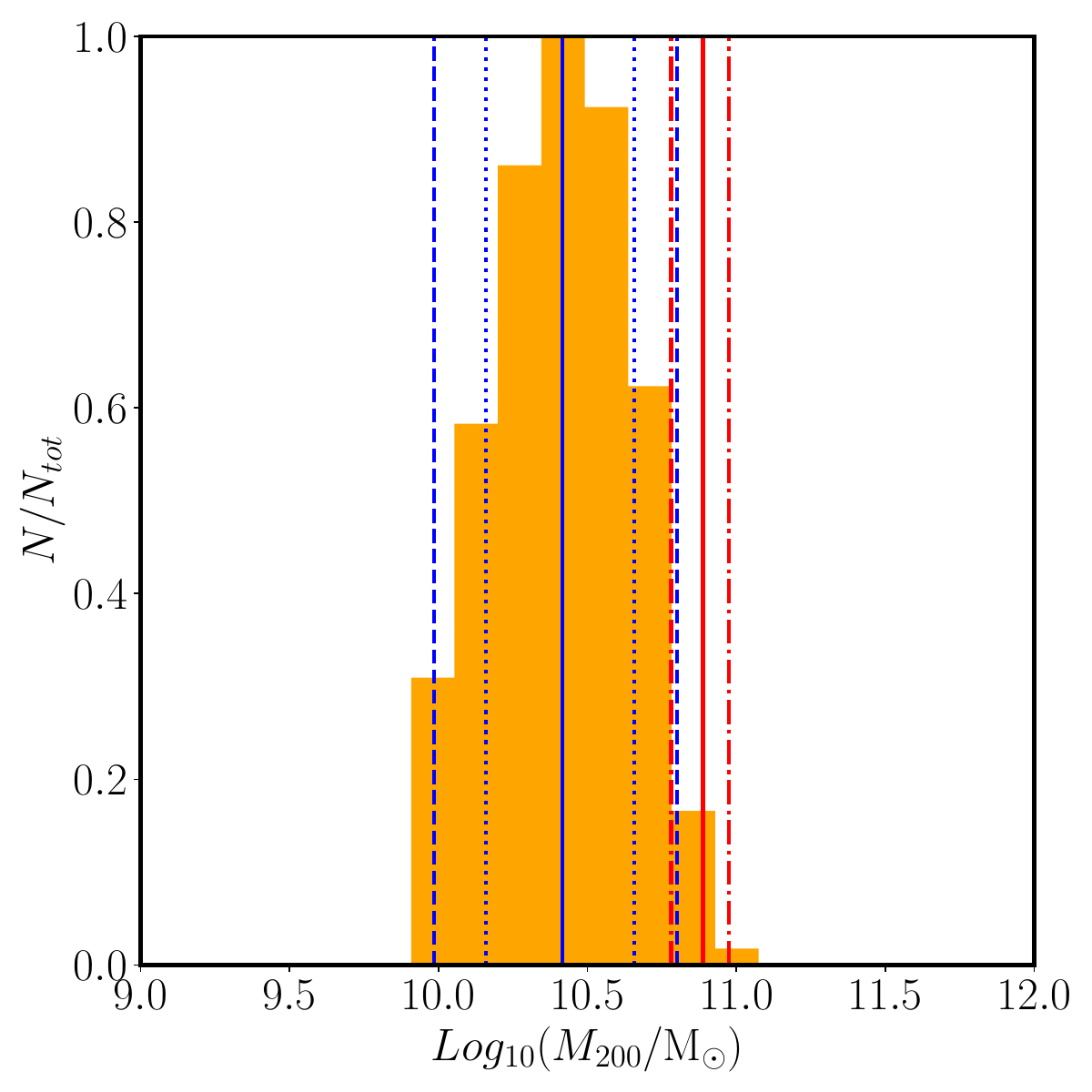}
\caption{Distribution of the $M_{200}$ values recovered by \textsc{GravSphere} (orange histogram). The mean value of the distribution is marked by the solid blue line, the dotted blue lines denote the 68\% confidence interval while the dashed blue lines indicate the 95\% confidence interval. The value reported by \citet{2019MNRAS.487.5799R} is marked by a solid red line with its 68\% confidence interval denoted by the dot-dashed red lines. \label{fig:m200_abund}}
\end{figure}

\textsc{GravSphere} also provides us with constraints on the halo virial mass, $M_{200} = 2.61_{-1.16}^{+1.95}\times 10^{10} M_{\odot}$ (see Fig.~\ref{fig:m200_abund}), and concentration parameter, $c_{200} = 16.58 \pm 11.27$.

Given the extensive tidal disruption experienced by the SMC, it is not entirely clear how we should interpret the recovered $M_{200}$ from present-day dynamical tracers. {\sc GravSphere} does attempt to model the impact of tidal stripping through the tidal radius and density fall off model parameters, $r_t$ and $\delta$ (see \S\ref{modeling} and Equation ~\ref{eqn:coreNFWtides}). Unfortunately, we could not obtain constraints on $r_t$ and $\delta$ that are bound only by our priors. Furthermore, {\sc GravSphere} is not able to account for historic mass loss from inside $r_t$, neither from tidal stripping nor tidal shocks \citep[e.g.][]{2006MNRAS.367..387R}. As such, any estimate of $M_{200}$ will be a {\it lower bound} on the SMC's pre-infall halo mass.

Despite the above caveats, {\sc GravSphere} yields an estimate of the SMC's pre-infall $M_{200}$ that overlaps, within our 95\% confidence intervals, with that obtained from abundance matching \citep[e.g.][]{2019MNRAS.487.5799R}: $M_{200,{\rm abund}} = 7.73\pm1.69\times 10^{10} M_{\odot}$ (see the solid and dashed red lines in Fig.~\ref{fig:m200_abund} that mark the median and $\pm 68\%$ confidence intervals of $M_{200,{\rm abund}}$).

Considering the, likely more robust, pre-infall $M_{200,{\rm abund}}$ estimation, our recovered $c_{200}$ parameter is consistent (within the 68\% uncertainty) with the value expected in $\Lambda$CDM \citep[$\sim$11 from][]{2014MNRAS.441.3359D} for a galaxy of the halo mass of the pre-infall SMC.

\subsubsection{The inner dark matter density of the SMC: testing dark matter heating models}\label{sec:densityprofile}

Armed with our recovered dark matter density profile and $M_{200}$ for the SMC, we now turn to its position in the $\rho_{\rm DM}$-$M_{200}$ plane. As first proposed in \citet{2019MNRAS.484.1401R} (and see also \S\ref{intro}), this provides a key test of \textquote{dark matter heating} models. For $M_{200}$, we will use the abundance matching pre-infall halo mass for the SMC: $M_{200,{\rm abund}} = 7.73\pm1.69\times 10^{10} M_{\odot}$ \citep{2019MNRAS.487.5799R}. As discussed above, this is likely to be a more reliable estimate than that based on the SMC's current dynamical state.

\begin{figure*}
\centering
     \begin{subfigure}[b]{0.48\textwidth}
         \centering
         \includegraphics[width=\textwidth, height=75mm]{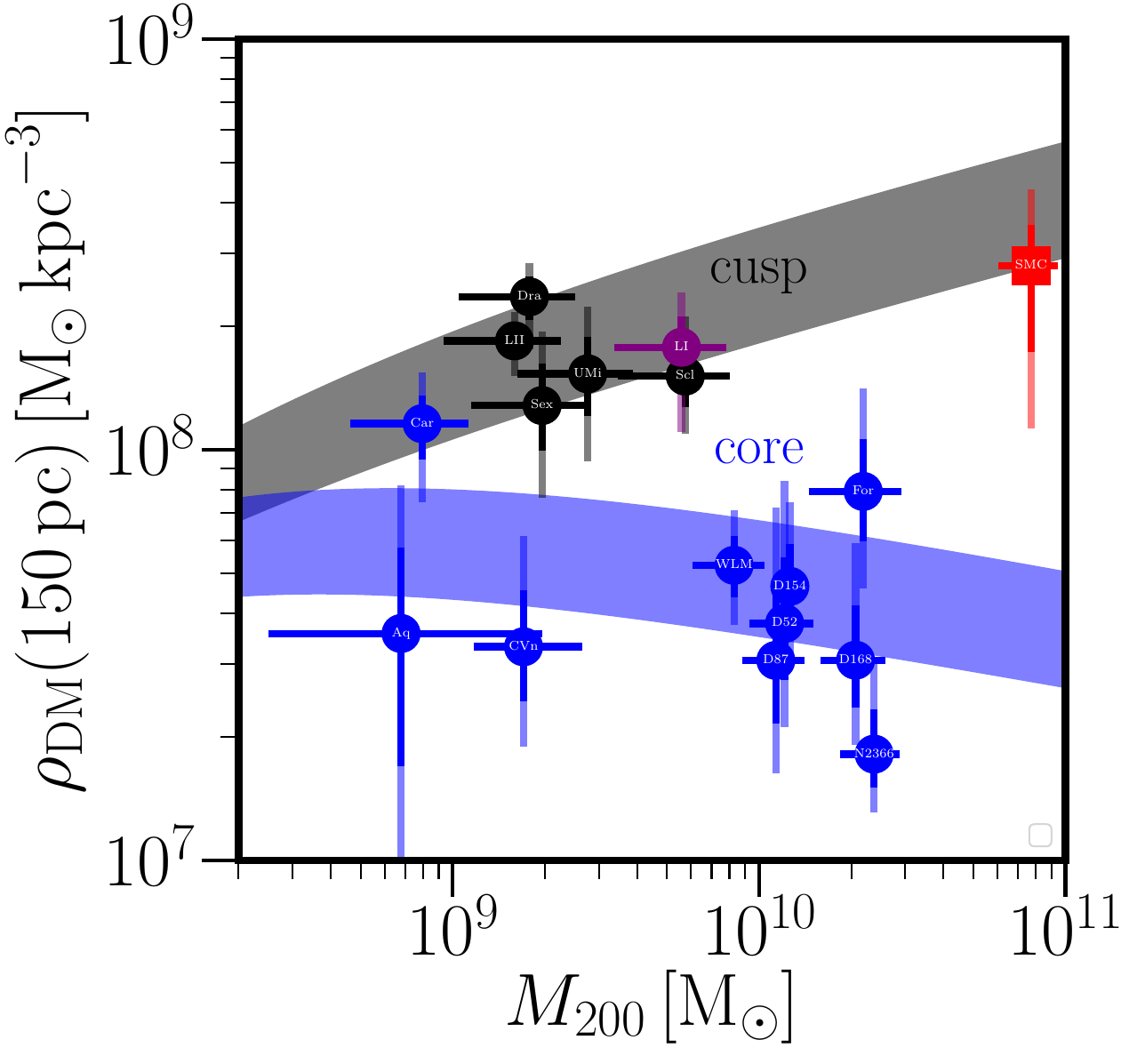}
     \end{subfigure}
     \begin{subfigure}[b]{0.48\textwidth}
         \centering
         \includegraphics[width=\textwidth, height=75mm]{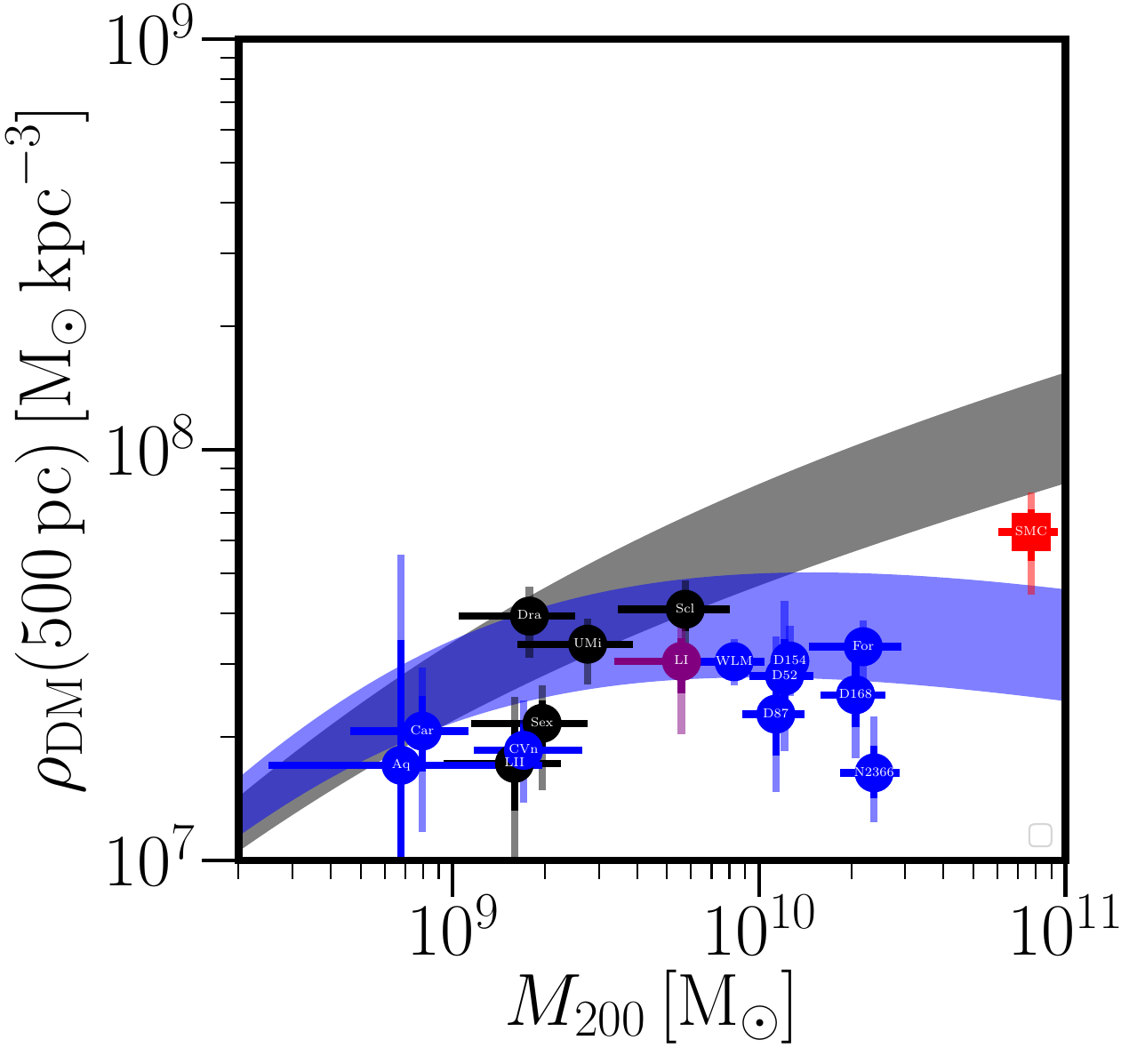}
     \end{subfigure}
    \caption{\textit{Left panel:} Inner dark matter density (at 150pc) as a function of halo mass for a sample of dwarf galaxies. The gray band marks the theoretical expectation for perfectly preserved cusps (no dark matter heating) while the light blue band denotes the expectation for complete core transformation (maximal dark matter heating). The black symbols are dwarf galaxies that stopped forming stars more than 6 Gyrs ago, the purple symbol is a galaxy that stopped forming stars between 6 and 3 Gyrs ago, the blue symbols are galaxies that stopped forming stars in the last 3 Gyrs and the red square is the SMC. The error bars for the symbols are fully coloured for the $1\sigma$ value and faintly coloured up to the $2\sigma$ value. \textit{Right panel:} same as the left panel but for the density estimated at 500pc (instead of 150pc). \label{fig:heating}}
\end{figure*}

Fig.~\ref{fig:heating} compares the theoretical expectations for perfectly preserved dark matter cusps in a $\Lambda$CDM cosmology (gray bands) and complete cusp-core transformations due to dark matter heating (light blue bands) with the data from \citep{2019MNRAS.484.1401R} (black, blue and purple circles) and the SMC (red square). The left panel shows estimates of the dark matter density at 150\,pc from the centres of the galaxies; the right panel at 500\,pc. The black symbols are galaxies that stopped forming stars more than 6\,Gyrs ago, the purple symbol is a galaxy that stopped forming stars  $3-6$ Gyrs ago and the blue symbols are galaxies that stopped forming stars in the last 3\,Gyrs. All galaxies have been selected to be tidally isolated today (see \citealt{2019MNRAS.484.1401R}).

Firstly, notice that at low $M_{200}$ the blue and black bands overlap. This is because the dark matter core size scales with $\sim R_{1/2}$ which in turn correlates with $M_{200}$. As $M_{200}$ is reduced, the expected core size shrinks and at a fixed length scale, the cusped and cored models begin to overlap. This happens at even higher mass for the $\rho_{\rm DM}(500\,{\rm pc})$ plot (right panel). Secondly, notice that the black and purple data points, corresponding to galaxies whose star formation shut down long ago, are consistent with dense dark matter cusps. By contrast, those dwarfs with recent star formation (blue data points) have had the most dark matter heating and are consistent with fully formed dark matter cores. The SMC, however, (red data point) has a much higher pre-infall $M_{200}$ than any of the data points taken from \citet{2019MNRAS.484.1401R}; at 150\,pc (left panel) it has a central density consistent with a dark matter cusp, albeit with large errors. At 500\,pc (right panel), despite the model being better-constrained (lower errors), it is not able to distinguish between a cusp or a core and both are, ever so slightly, consistent within the errors. The SMC conserving its dark matter cusp is in line with the evidence of cusps being resilient to tides \citep{2020MNRAS.491.4591E,2023MNRAS.519..384E}.

\begin{figure*}
     \centering
     \begin{subfigure}[b]{0.48\textwidth}
         \centering
         \includegraphics[width=\columnwidth,height=90mm]{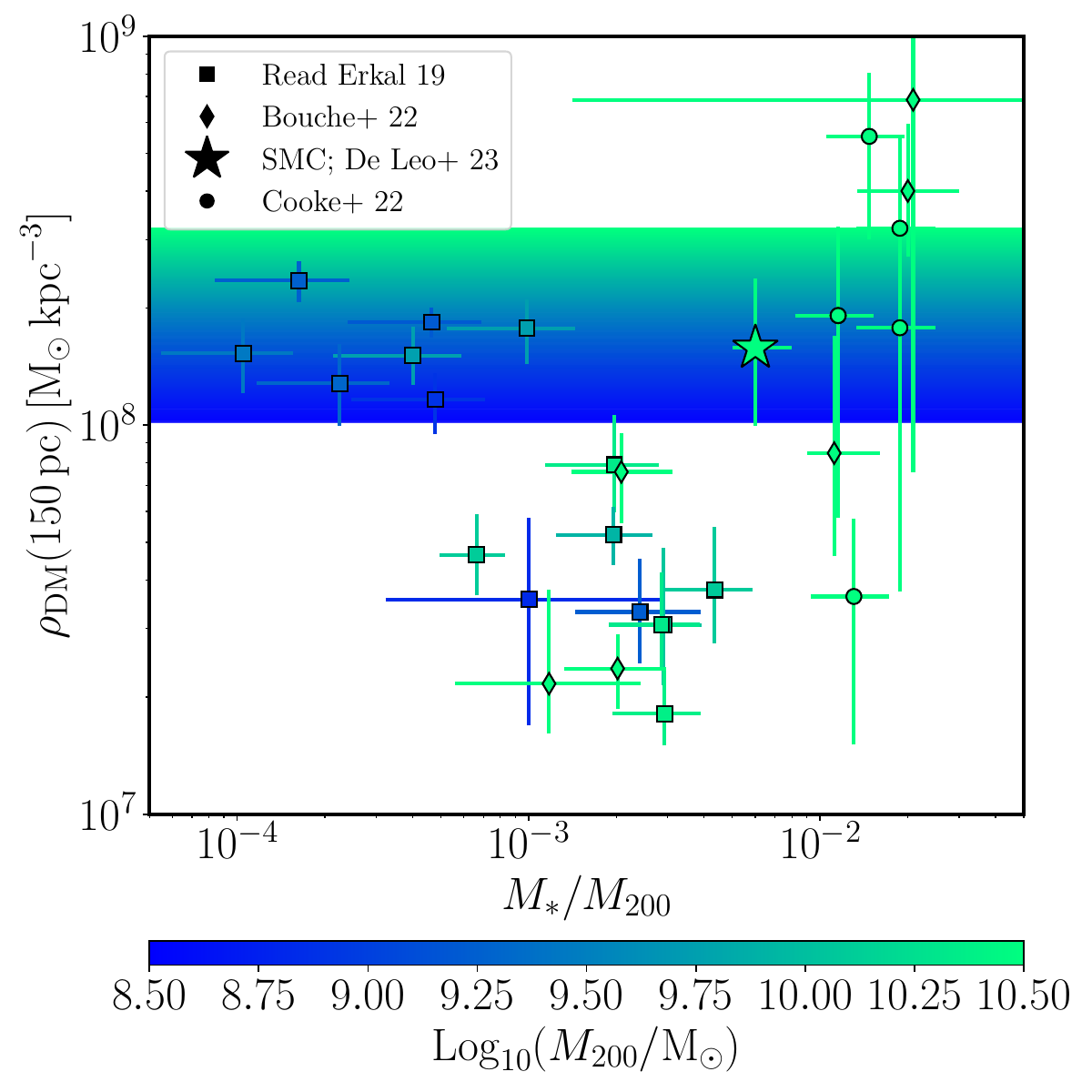}
     \end{subfigure}
     \begin{subfigure}[b]{0.48\textwidth}
         \centering
         \includegraphics[width=\columnwidth,height=90mm]{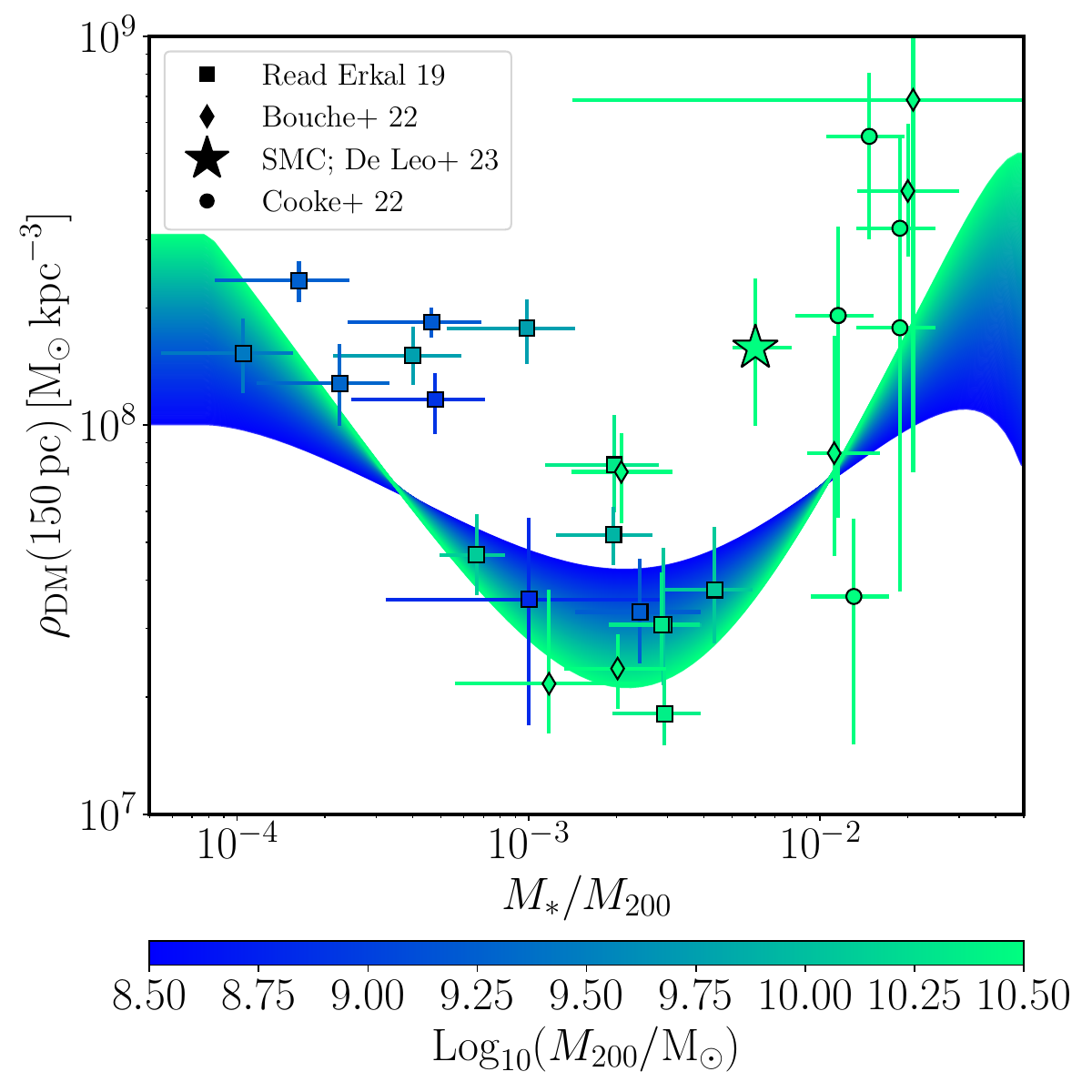}
     \end{subfigure}
     \newline
     \centering
     \begin{subfigure}[b]{0.48\textwidth}
         \centering
         \includegraphics[width=\textwidth,height=90mm]{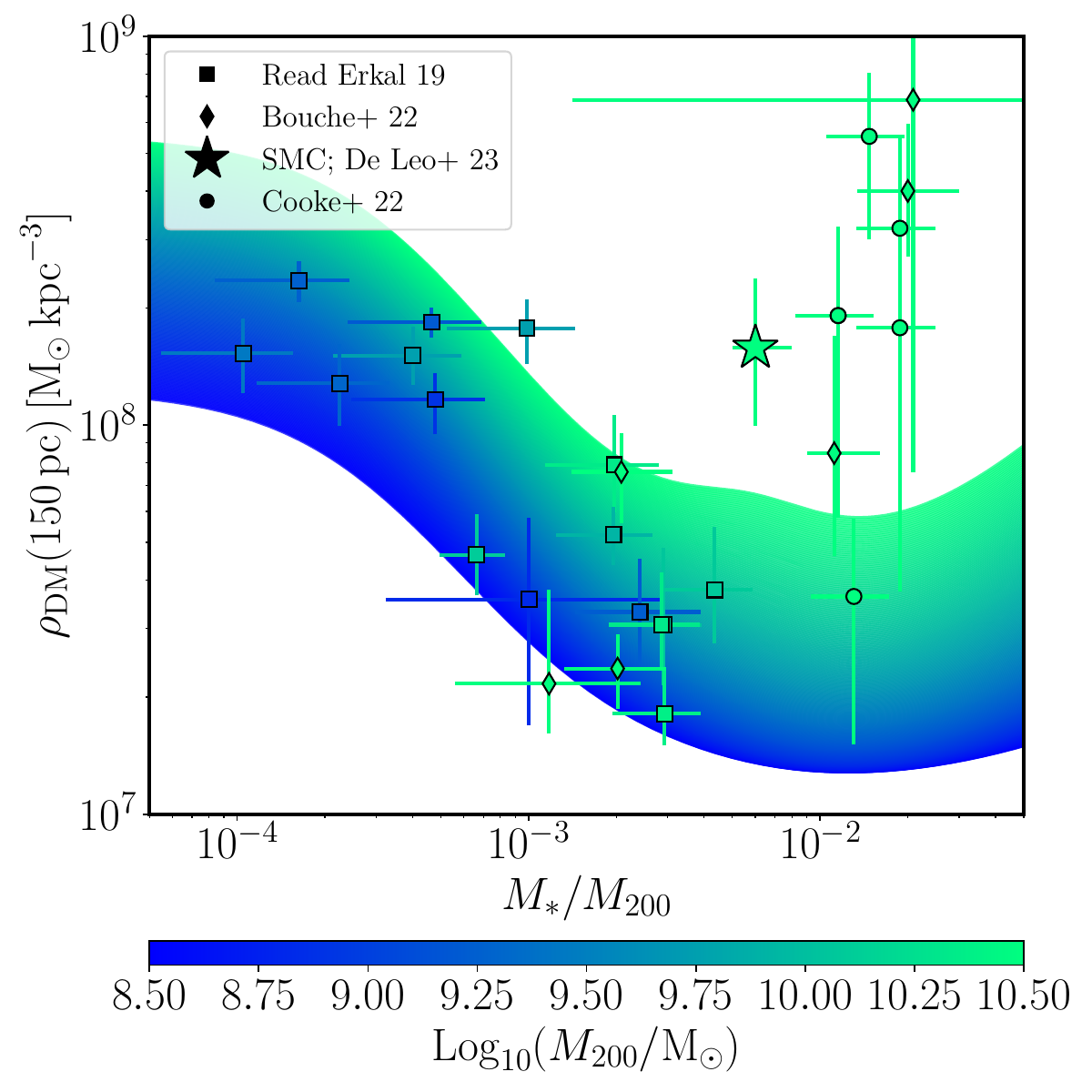}
     \end{subfigure}
     \begin{subfigure}[b]{0.48\textwidth}
         \centering
         \includegraphics[width=\textwidth,height=90mm]{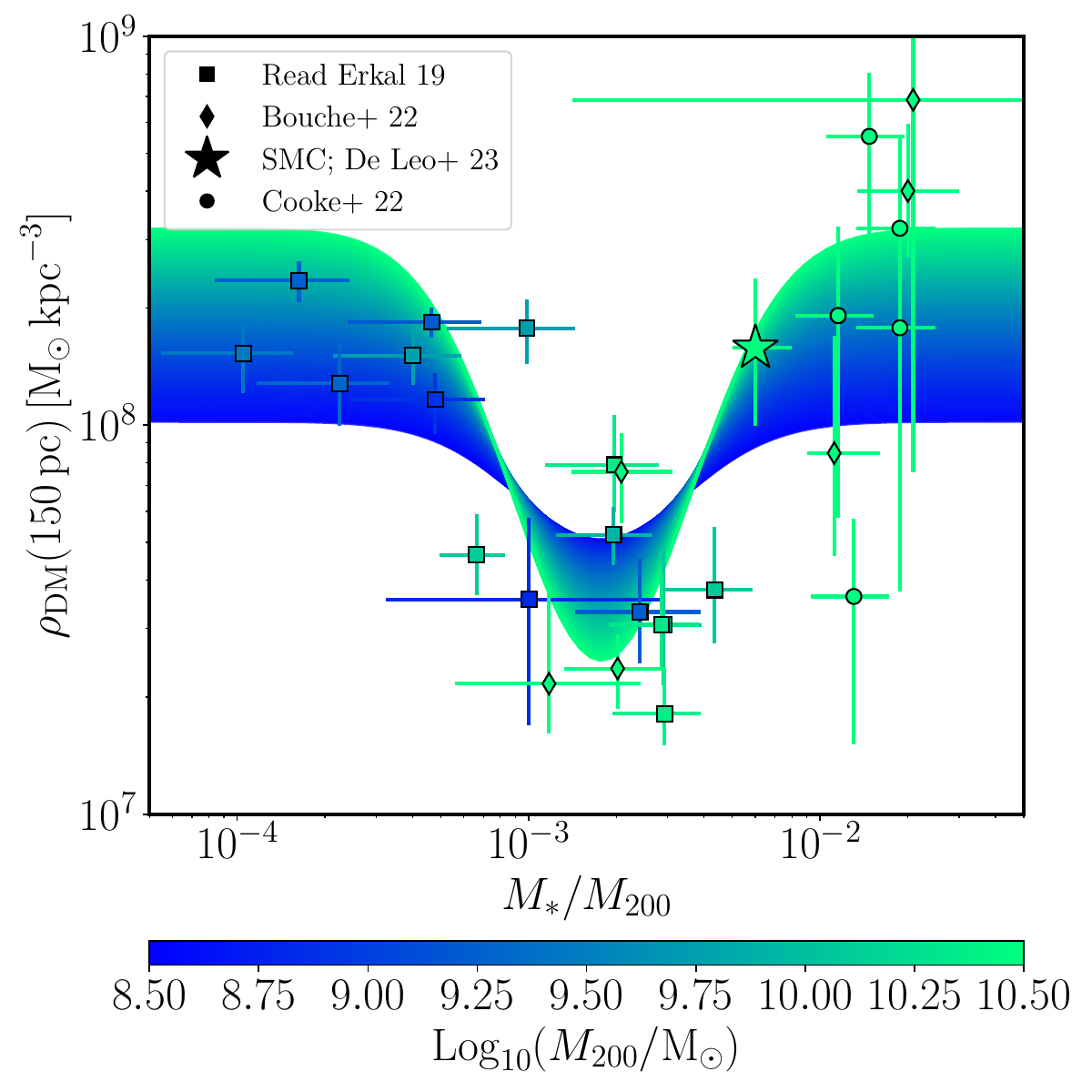}
     \end{subfigure}
        \caption{Central dark matter density $\rho_{DM}$ at 150\,pc from the galactic centre against stellar-to-halo mass ratio $M_{*}/M_{200}$ for the data presented in \citet{2019MNRAS.487.5799R}, \citet{2022A&A...658A..76B}, \citet{2022MNRAS.512.1012C} and the SMC (the coloured symbols) compared to different dark matter models (coloured bands). In all panels, the colour of the points and of specific positions along the bands marks the pre-infall halo mass, $M_{200}$ (see the colourbar). \textit{Top left panel:} the band is the prediction of the \citet{1996ApJ...462..563N} model. \textit{Top right panel:} the band is the prediction of the \citet{2014MNRAS.441.2986D} model. \textit{Bottom left panel:} the band is the prediction of the \citet{2020MNRAS.497.2393L} model. \textit{Bottom right panel:} The band is a modified coreNFW model tuned to the data.}
        \label{fig:multi_prof}
\end{figure*}

We now consider whether the above behaviour -- galaxies moving from being cusped to cored and then back to cusped again -- is consistent with dark matter heating models. To test this, we switch from the $\rho_{\rm DM}$-$M_{200}$ plane to the $\rho_{\rm DM}$-$M_*/M_{200}$ plane. As discussed in \S\ref{intro}, $M_*/M_{200}$ -- to leading order -- indicates how much energy is available to drive dark matter heating \citep[e.g.][]{2012ApJ...759L..42P}. We expect dark matter heating to increase with increasing $M_*/M_{200}$ until the self-gravity of the stars begins to dominate over the dark matter at which point dark matter heating becomes inefficient again \citep[e.g.][]{2014MNRAS.437..415D}. In Fig.~\ref{fig:multi_prof}, we combine our data for the SMC with literature data from \citealt{2019MNRAS.484.1401R}, \citealt{2022A&A...658A..76B} (courtesy of N. F. Bouch{\'e}) and \citealt{2022MNRAS.512.1012C} (courtesy of R. C. Levy) to explicitly test this. We can see from Fig.~\ref{fig:multi_prof} the relationship between central dark matter density at 150\,pc and the stellar-to-halo mass ratio, $M_*/M_{200}$, for the data (squares and stars) as compared to several different models (coloured lines). The colours denote tracks of constant $M_{200}$, as marked by the colourbars. The data points are coloured similarly by their median $M_{200}$, as estimated from abundance matching.
The top left panel of Fig.~\ref{fig:multi_prof} shows a classical NFW model \citep{1996ApJ...462..563N} without dark matter heating. This model fits the more dense `cusp'-like dwarfs, but fails to reproduce the lower density `core'-like dwarfs in the range: $5 \times 10^{-4} \lesssim M_{*}/M_{200} \lesssim 10^{-2}$.
The top right panel of Fig.~\ref{fig:multi_prof} shows the \citet{2014MNRAS.441.2986D} model which correctly reproduces the qualitative behaviour seen in the data, with cusp-like densities below $M_*/M_{200} \lesssim 5 \times 10^{-4}$, core-like densities in the range $5 \times 10^{-4} \lesssim M_*/M_{200} \lesssim 5 \times 10^{-3}$ and cusp-like densities again for $M_*/M_{200} \gtrsim 5 \times 10^{-3}$. However, there are quantitative differences, with the model favouring a slower and smoother transition between the cusped and cored regimes (and back again) as compared to the data (for example, cusp-like densities at $M_*/M_{200} \sim 5 \times 10^{-4}$ are not expected by the model). We must note that the model has been computed assuming the {\it median} concentration parameter, $c_{200}$, in a $\Lambda$CDM cosmology \citep{2014MNRAS.441.3359D}. This may not be appropriate for the dwarf spheroidal satellites of the Milky Way that likely fell in long ago \citep{2019MNRAS.487.5799R} and may, therefore, be biased to higher concentration parameters \citep[e.g.][]{2008MNRAS.391.1685S}. In  Appendix~\ref{app_dicintio} we explore the effect of the introduction of a $2-\sigma$ bias above the median in the estimation of the concentration parameter of dark matter halos, $c_{200}$. Including this bias, the model does pass comfortably through the data points for the high central density dwarf spheroidals at low and high $M_*/M_{200}$ (including the SMC). Whether this is the correct interpretation of the behaviour of these data, or whether the dark matter heating model of \citet{2014MNRAS.437..415D} is not quite correct, remains to be seen.

The bottom left panel of Fig.~\ref{fig:multi_prof} shows the \citet{2020MNRAS.497.2393L} model which correctly reproduces the behaviour seen in the data up until $M_*/M_{200} \sim 10^{-2}$, but fails to account for the more dense halos at higher mass ratios. The errors for most of these higher mass ratio data points remain large, but the data point we derive here for the SMC certainly seems to be in significant tension with the predictions from \citet{2020MNRAS.497.2393L}. This highlights two important points: (i) not all dark matter heating models in the literature make the same predictions, and (ii) the latest data are now able to quantitatively test these models.

Finally, in the bottom right panel of Fig.~\ref{fig:multi_prof} we show a handy analytic function, built on the {\sc coreNFW} profile (equation \ref{eqn:coreNFW}), that captures the main features of the data. This introduces an $M_*/M_{200}$ dependence on the $n$ parameter (that determines how cusped or cored the profile is):

\begin{equation}
n = \kappa_3\exp\left(\frac{-(\rm{Log}_{10}(M_*/M_{200})-\kappa_1)^2}{\kappa_2}\right)
\end{equation}
where $\kappa_1 = -2.75$, $\kappa_2 = 0.2$ and $\kappa_3 = 1.25$. Readers may find this useful as a compact analytic description of the behaviour of the data and/or to test their own favoured models.

\subsubsection{The astrophysical $J$-factor and $D$-factor of the SMC}

\begin{figure}
\centering
\includegraphics[width=\columnwidth]{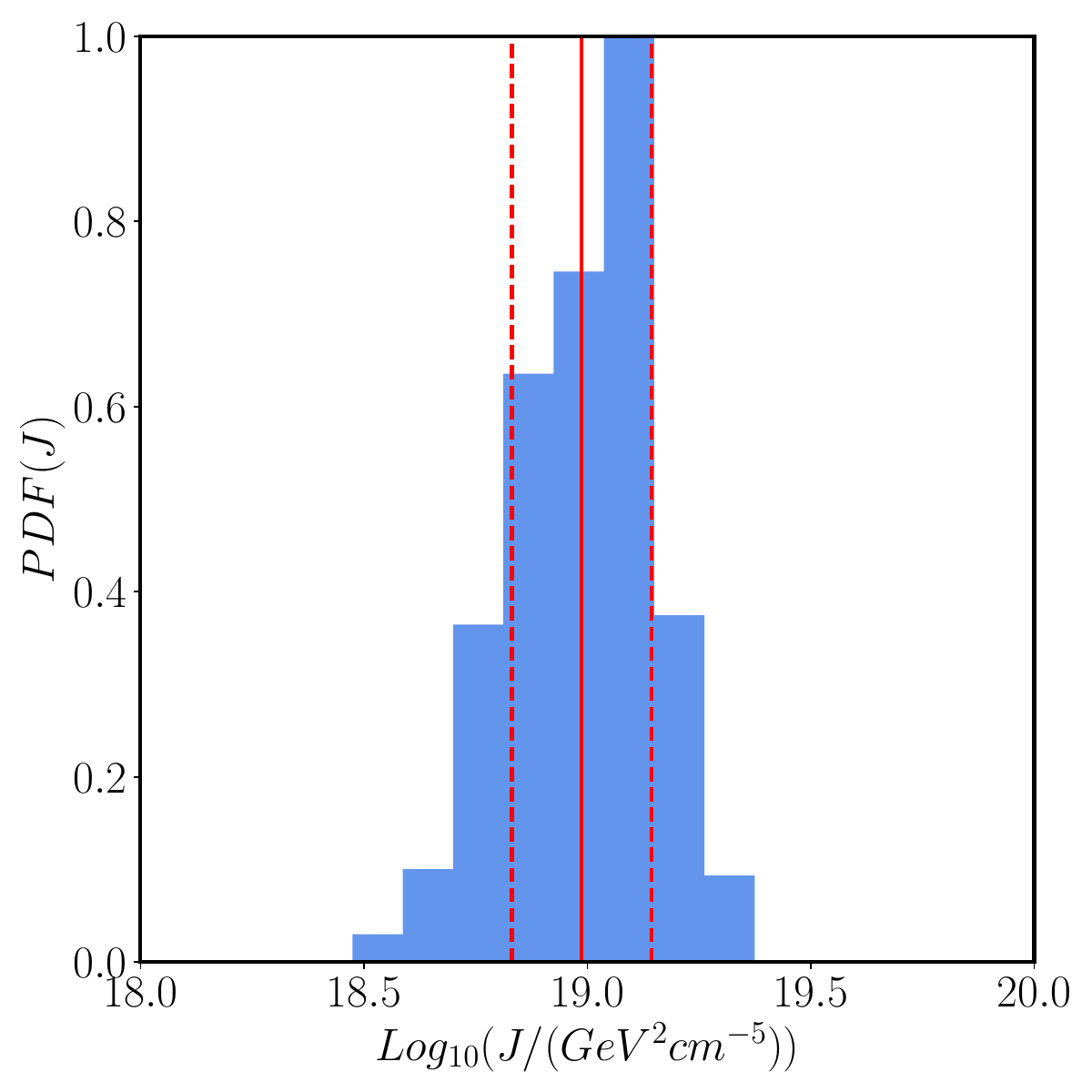}
\caption{The PDF for the $J$-factor recovered by \textsc{GravSphere} (blue histogram). The mean value of the distribution is marked as a solid red line with dashed red lines marking the 68\% confidence interval. \label{fig:jay}}
\end{figure}

Given their dense environments, dwarf galaxies can be suitable candidates for searches of dark matter annihilation and/or decay events \citep{2010AdAst2010E..45K} so we will conclude this section with a look at the SMC in this context. The density estimation of \textsc{GravSphere} can be used to derive the $J$-factor: the integral of the square of the dark matter density along the line-of-sight and over a solid angle $\delta\Omega$ \citep{2020JCAP...09..004A}. This parameter quantifies the dependence of dark matter annihilation searches on the density of the astrophysical target being searched. We recovered the distribution of $J$-factors for the SMC, shown in Fig.~\ref{fig:jay}. This has a mean of $18.99\pm0.16$\, GeV$^2$\,cm$^{-5}$, shown as the solid red line in the figure. We also recovered the distribution of the $D$-factor: the integral of the dark matter density along the line-of-sight and over a solid angle $\delta\Omega$ \citep{2020JCAP...09..004A}. This is the relevant quantity for testing decaying dark matter models. We find a mean value of $18.73\pm0.04$\, GeV$^2$\,cm$^{-5}$. Both the $J$- and $D$-factor are interestingly competitive with the densest dwarf spehroidals known to date around the Milky Way \citep[e.g.][]{2020JCAP...09..004A}, suggesting that the SMC is a prime target for such annihilation and decay searches. While the SMC has a much stronger gamma-ray background than other nearby gas-free dwarf spheroidal galaxies (like Draco), this background can be well-modelled using the observed distribution of stellar light, gas and high energy point sources, yielding competitive constraints on dark matter models \citep[i.e.][]{2016PhRvD..93f2004C}.

\section{Discussion}\label{discussion}

\subsection{The impact of priors}\label{sec:disc_priors}
Before delving deeper into the information that can be extracted from the mass model of the SMC, it is worth discussing briefly the choices of priors operated throughout the modeling process and how they affect the results.
The flat priors assumed for the mass profile (rows 1 to 6 in Tab.~\ref{tab:parampriors}) were purposefully weak to allow for the recovery of any kind of final model (be it cuspy or cored). The $M_{200}$ and $c_{200}$ bounds were informed by previous studies of the SMC (respectively \citealt{2019MNRAS.487.5799R} and \citealt{2012MNRAS.421.2109B}) while the others were left wide to allow for any possible solution.
The priors assumed for the velocity PDFs recovered by the \textsc{binulator} (rows 13 to 15 of Tab.~\ref{tab:parampriors}) were informed by our previous study of the SMC bulk motion \citep{2020MNRAS.495...98D}, based on the same observational data. For the proper decontamination of the debris it was also fundamental to allow the \textsc{binulator} to use at least two secondary Gaussians (priors on rows 16 to 21 of Tab.~\ref{tab:parampriors}) to fit for contaminants on either side of the SMC bulk velocity distribution. As for the priors on the mass profile, we favoured weaker priors for the anisotropy parameters (rows 9 to 12 of Tab.~\ref{tab:parampriors}). We found that tighter priors (i.e. $-0.01 \leq \beta_0 \leq 0.01$) produced a slightly more cored density profile at the expense of strongly enforcing a zero value of the central anisotropy profile (see Fig.~\ref{fig:tight_beta_prior} in Appendix~\ref{app_priors}).

\subsection{The present-day mass of the SMC}\label{sec:discussionmass}

It is difficult to make a proper comparison of our recovered present-day mass of the SMC with values in the literature as most estimations were derived from mass models that assumed gas and stars were bound by the SMC potential to large radii. We obtain $M_{DM}(\leq 3\,{\rm kpc})=1.39^{+0.08}_{-0.07} \times 10^9 M_{\odot}$ and $M_{*}(\leq 3\,{\rm kpc})=0.34\pm0.05 \times 10^9 M_{\odot}$ (\S\ref{sec:SMCmasS}). Summing to this the total gas mass measured within the same radius ($5.6 \times 10^8 M_{\odot}$; \citealt[ ][]{1999MNRAS.302..417S,2004ApJ...604..176S,2005A&A...432...45B}), we obtain a total present-day mass of the SMC equal to $M_{{\rm tot}}(\leq 3\,{\rm kpc})=2.29\pm0.46 \times 10^9 M_{\odot}$.
While this value is consistent with the estimate for total SMC mass of $2.4 \times 10^9 M_{\odot}$ in \citet{2004ApJ...604..176S}, the underlying assumptions of our methods are quite different (the model in \citealt{2004ApJ...604..176S} was a two-component model without dark matter) so it is challenging to meaningfully compare the two values. Our total mass is also consistent with the lower bound of the estimation from \citet{2006AJ....131.2514H}, who derived a total mass between $2.7 \times 10^9 M_{\odot}$ and $5.1 \times 10^9 M_{\odot}$ through a simple virial analysis based on stellar kinematics. The smaller value that we recover is due to the fact our model excludes the stars in the tidal debris from the computation of the bound SMC mass.

\subsection{Comparison with dark matter annihilation literature} \label{sec:discussionJD}

Our recovered value for the $J$-factor ($18.99\pm0.16$\, GeV$^2$\,cm$^{-5}$) is in good agreement with the estimate of \citet{2016PhRvD..93f2004C} and -- interestingly -- on par with the isolated dwarf galaxy Draco \citep[$18.69\pm0.05$\, GeV$^2$\,cm$^{-5}$ estimated in ][]{2020JCAP...09..004A}. The estimated $D$-factor ($18.73\pm0.04$\, GeV$^2$\,cm$^{-5}$) likewise is consistent with estimations for isolated dwarf galaxies \citep[Draco, Tucana II estimated in ][]{2016PhRvD..93j3512E}. This suggests that the SMC is a competitive target for the observation of gamma-rays and/or X-rays originating from dark matter annihilation and/or decay events, provided that its gamma-ray background is correctly modelled.

\section{Conclusions}\label{conclusions}

Using mock data we showed that, despite being subjected to heavy tidal disruption, the SMC can still be mass modelled with methods that require dynamical equilibrium. For this, we assumed that the galaxy is composed of a central bound remnant surrounded by tidal debris (as supported by the latest observational data, e.g. \citealt{2020ApJ...904...13G}). Given that building an unbiased mass model requires careful removal of tidal debris along the line-of-sight, we introduced the \textsc{binulator}. We showed that this new method to achieve the decontamination successfully worked on mock data for a tidally disrupting SMC.

We then proceeded to apply a Jeans mass modelling
method (\textsc{binulator+GravSphere}) to ${\sim}6000$ RGB stars with spectroscopic and proper motion data from \textit{Gaia} EDR3 to
build a new mass model of the SMC. The data decontamination employed by the \textsc{binulator} and the use of the full dynamical information (the line-of-sight velocity distribution and proper motions) by \textsc{GravSphere} were instrumental in recovering a robust model which we could use to further explore the characteristics of the dark matter halo of the SMC. After the removal of the tidally unbound interlopers, we recovered both the mass density and the stellar velocity anisotropy profile (which shows the remaining stars to be isotropic at all radii within the uncertainties).

We provided a new estimate for the total present-day mass of the SMC, $M_{{\rm tot}}(\leq 3\,{\rm kpc})=2.29\pm0.46 \times 10^9 M_{\odot}$, based on stellar kinematics, which takes into account the extensive tidal disruption undergone by the galaxy.

Our model found that the SMC has a high central density, $\rho_{\rm DM}(150\,{\rm pc}) = 1.58_{-0.58}^{+0.80}\times 10^8 M_{\odot} \rm kpc^{-3}$, which is consistent with a dark matter cusp within the $\Lambda$CDM paradigm. The inferred dark matter density profile provides an observational reference point for the halo mass scale at which dark matter heating becomes inefficient and is no longer able to drive a cusp-core transformation.

We used the SMC, together with previously available data, to test dark matter heating models in the literature, finding good qualitative agreement with the \citet{2014MNRAS.437..415D} model but poorer agreement with the \citet{2020MNRAS.497.2393L} model at $M_*/M_{200} > 10^{-2}$. We also introduced a new analytic density profile that gives a good fit to the central dark matter density of dwarf galaxies and its dependence on $M_*/M_{200}$.

Finally, from the recovered cuspy dark matter density profile, we derived an astrophysical $J$-factor of $18.99\pm0.16$\, GeV$^2$\,cm$^{-5}$ ($D$-factor of $18.73\pm0.04$\, GeV$^2$\,cm$^{-5}$), suggesting that the SMC is a very promising target for dark matter annihilation and decay searches.

\section*{Acknowledgements}

The authors thank the anonymous referee for their insightful comments that improved the paper. MDL thanks Alessia Gualandris, Jorge Pe\~{n}arrubia and Alex Drlica-Wagner for insightful comments and discussions which helped improve the present work. MDL also thanks Nicolas F. Bouch\'{e} and Rebecca C. Levy for providing access to their data. The research leading to these results has received funding from the European Community's Seventh Framework Programme (FP7/2013-2016) under grant agreement number 312430 (OPTICON). This work was also funded by ANID, Millenium Science Initiative, ICN12\_009.

\textit{Software:} this research made use of the Astropy \citep{2013A&A...558A..33A, 2018AJ....156..123A}, Healpy \citep{2005ApJ...622..759G,Zonca2019}, Matplotlib \citep{2007CSE.....9...90H} and Numpy \citep{harris2020array} packages.

\section*{Data Availability Statement}

The data underlying this article will be shared on reasonable request to the corresponding author.

\bibliographystyle{mnras}
\bibliography{surviving_SMC}

\appendix

\section{\textsc{GravSphere} recovered profiles}\label{app_profiles}

\begin{figure*}
     \centering
     \begin{subfigure}[b]{0.45\textwidth}
         \centering
         \includegraphics[width=\textwidth, height=70mm]{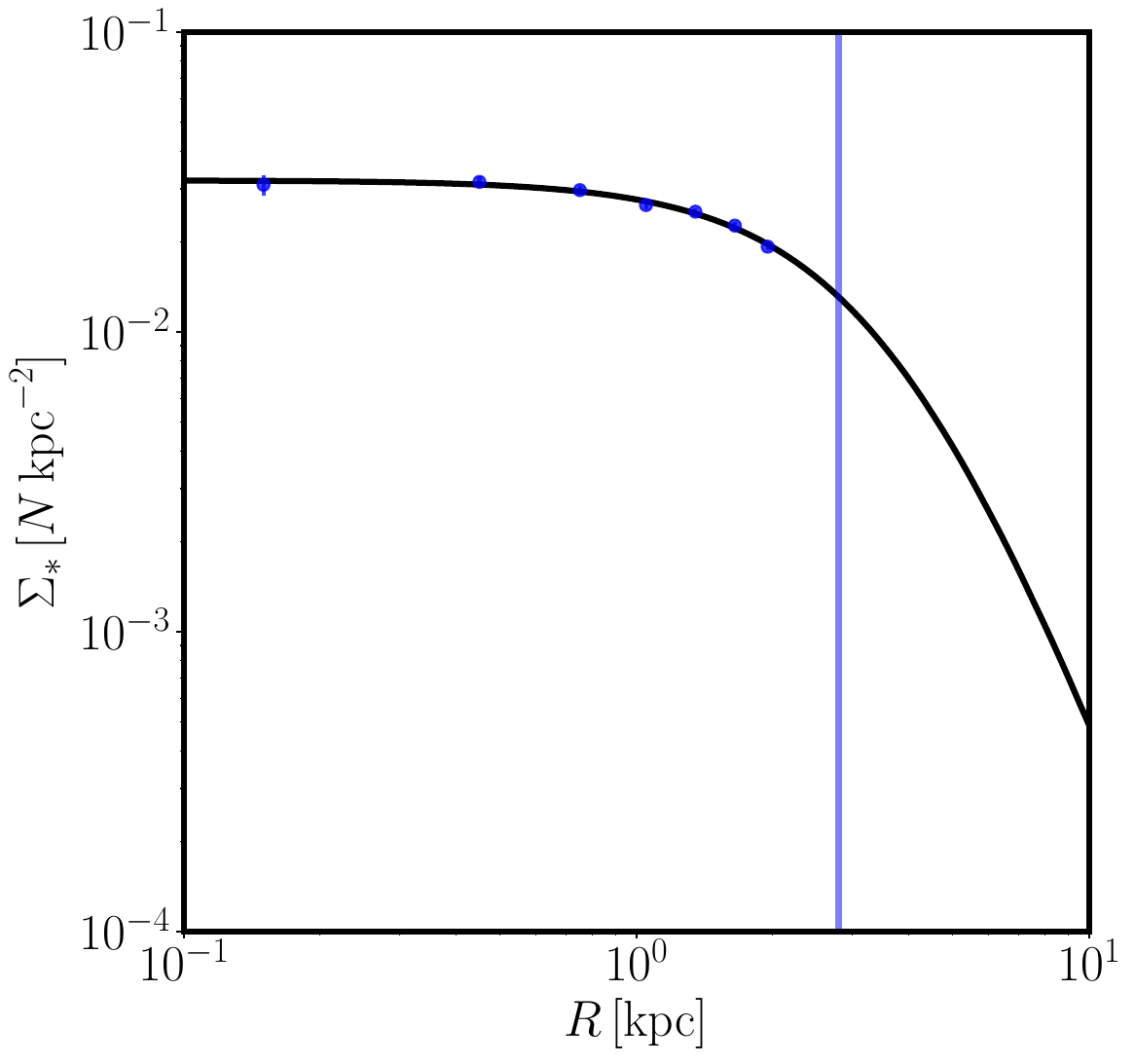}
     \end{subfigure}
     \begin{subfigure}[b]{0.45\textwidth}
         \centering
         \includegraphics[width=\textwidth, height=70mm]{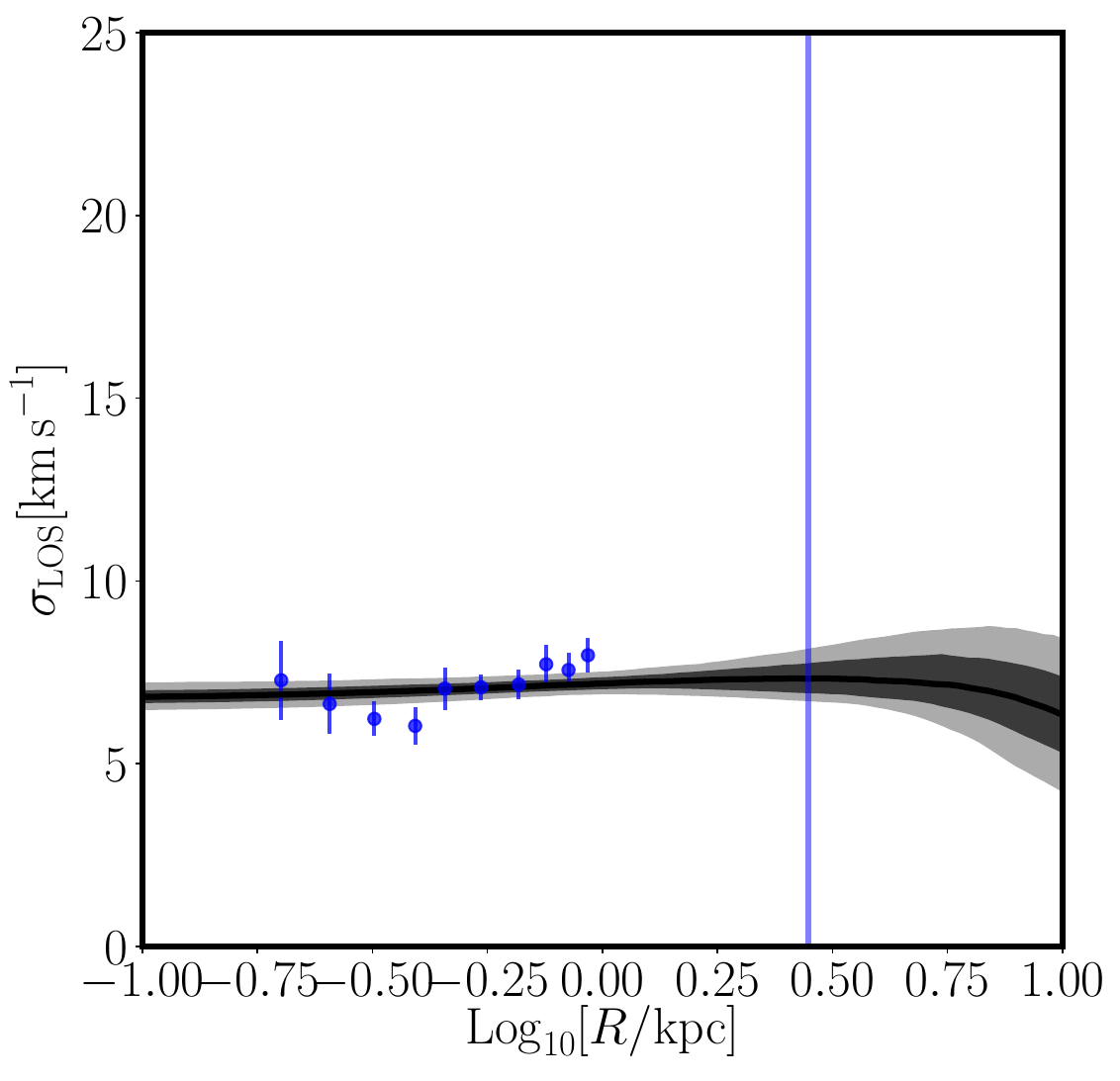}
     \end{subfigure}\\
	\centering
     \begin{subfigure}[b]{0.45\textwidth}
         \centering
         \includegraphics[width=\textwidth, height=70mm]{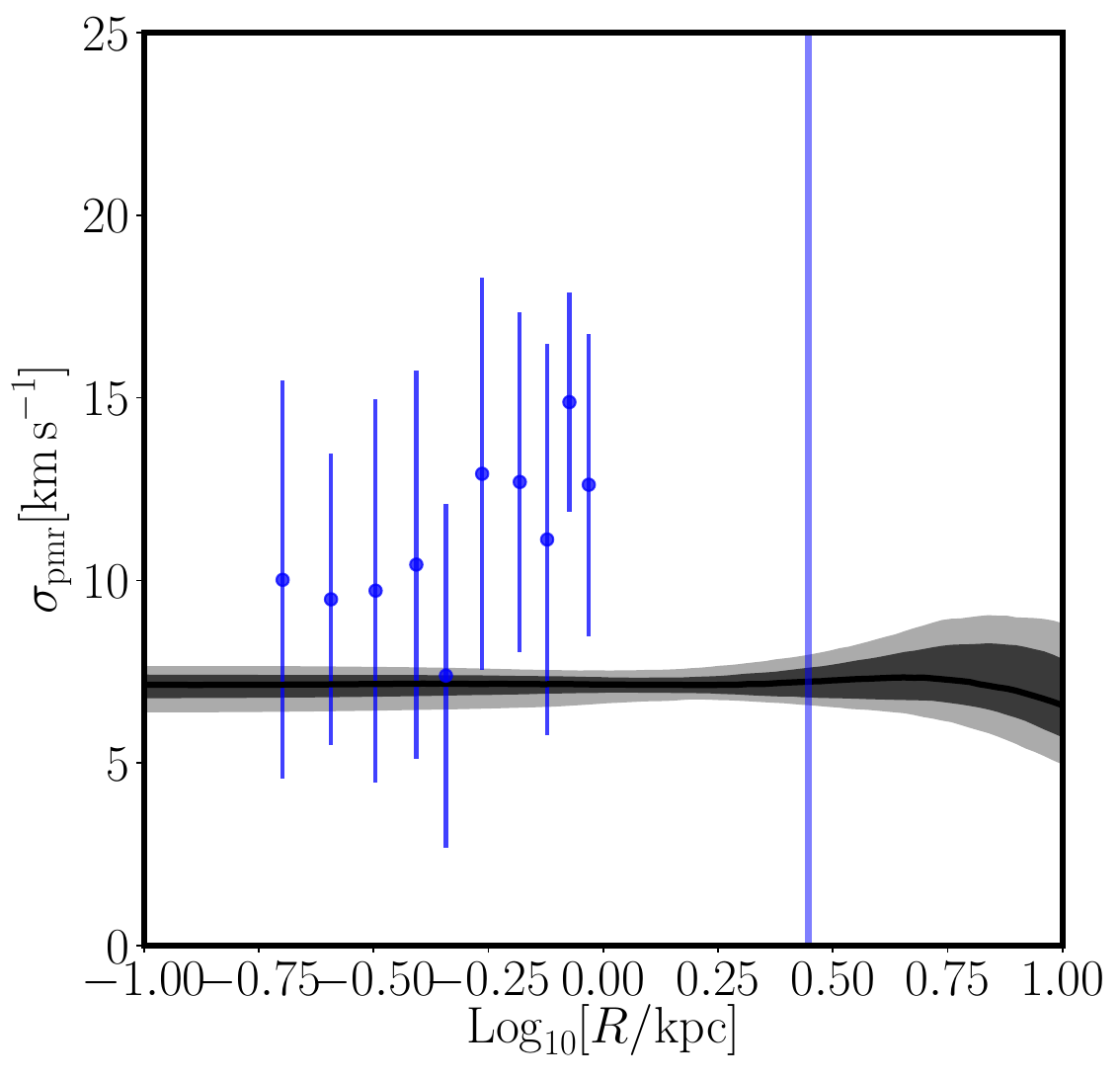}
     \end{subfigure}
     \begin{subfigure}[b]{0.45\textwidth}
         \centering
         \includegraphics[width=\textwidth, height=70mm]{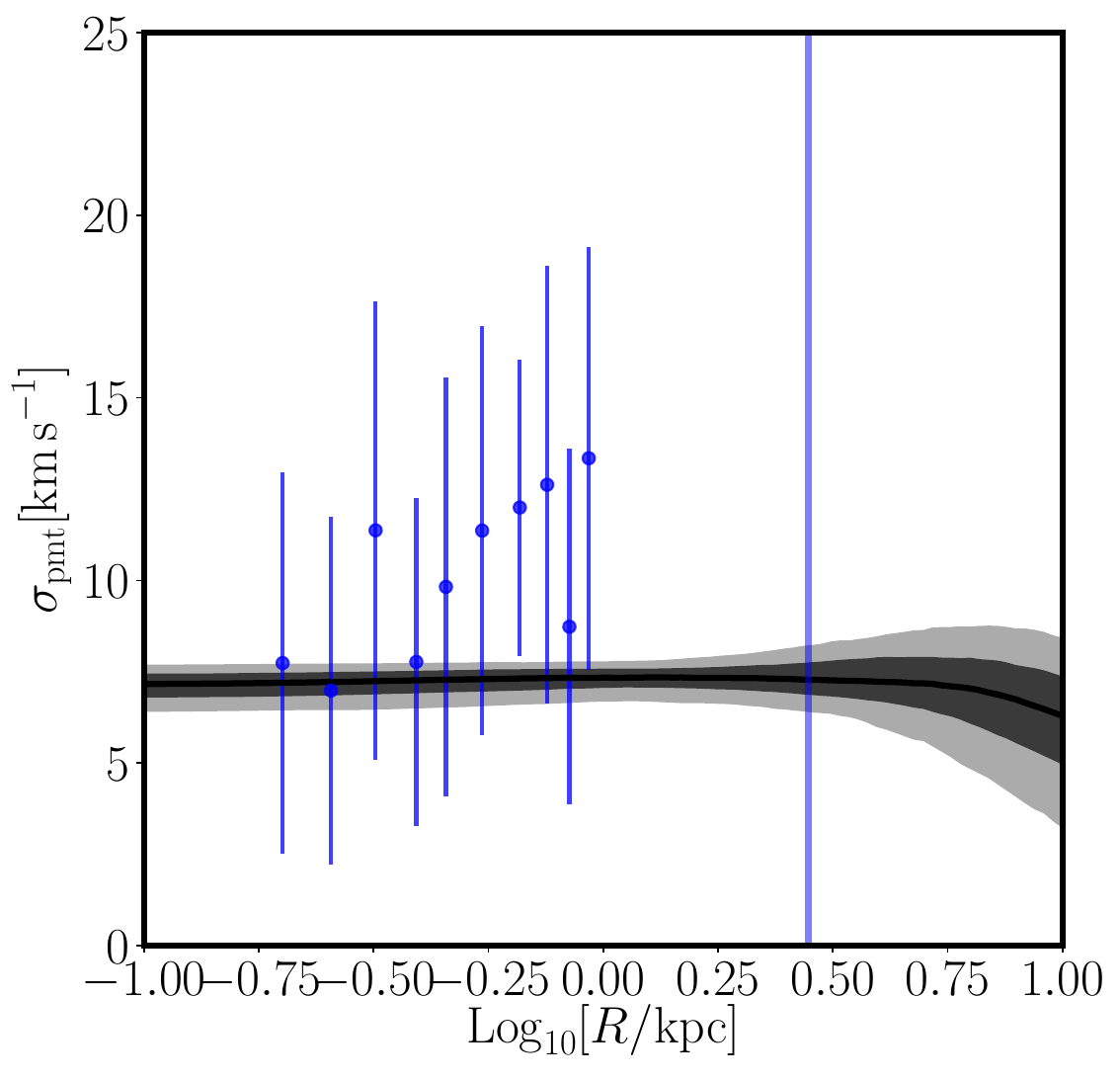}
     \end{subfigure}
        \caption{\textsc{GravSphere} recovered profiles for the heavily disrupted simulation. \textit{Top left panel}: surface brightness $\Sigma_*(r)$. The black line is the \textsc{GravSphere} fit, the blue points are the simulation data and the faint blue line marks the half-light radius computed by \textsc{GravSphere}. \textit{Top right panel}: velocity dispersion profile along the line of sight $\sigma_{LOS}$, the black line is \textsc{GravSphere} best fit solution, the dark and light grey contour are respectively the 68\% and the 95\% confidence intervals, the blue points with error bars are the binned data and the faint blue line marks the half-light radius computed by \textsc{GravSphere}. \textit{Bottom left panel}: radial velocity dispersion profile $\sigma_{rad}$, lines and symbols are the same as for $\sigma_{LOS}$. \textit{Bottom right panel}: tangential velocity dispersion profile $\sigma_{tan}$, lines and symbols are the same as for $\sigma_{LOS}$.}
        \label{fig:disrupted_profs}
\end{figure*}

\begin{figure*}
     \centering
     \begin{subfigure}[b]{0.45\textwidth}
         \centering
         \includegraphics[width=\textwidth, height=70mm]{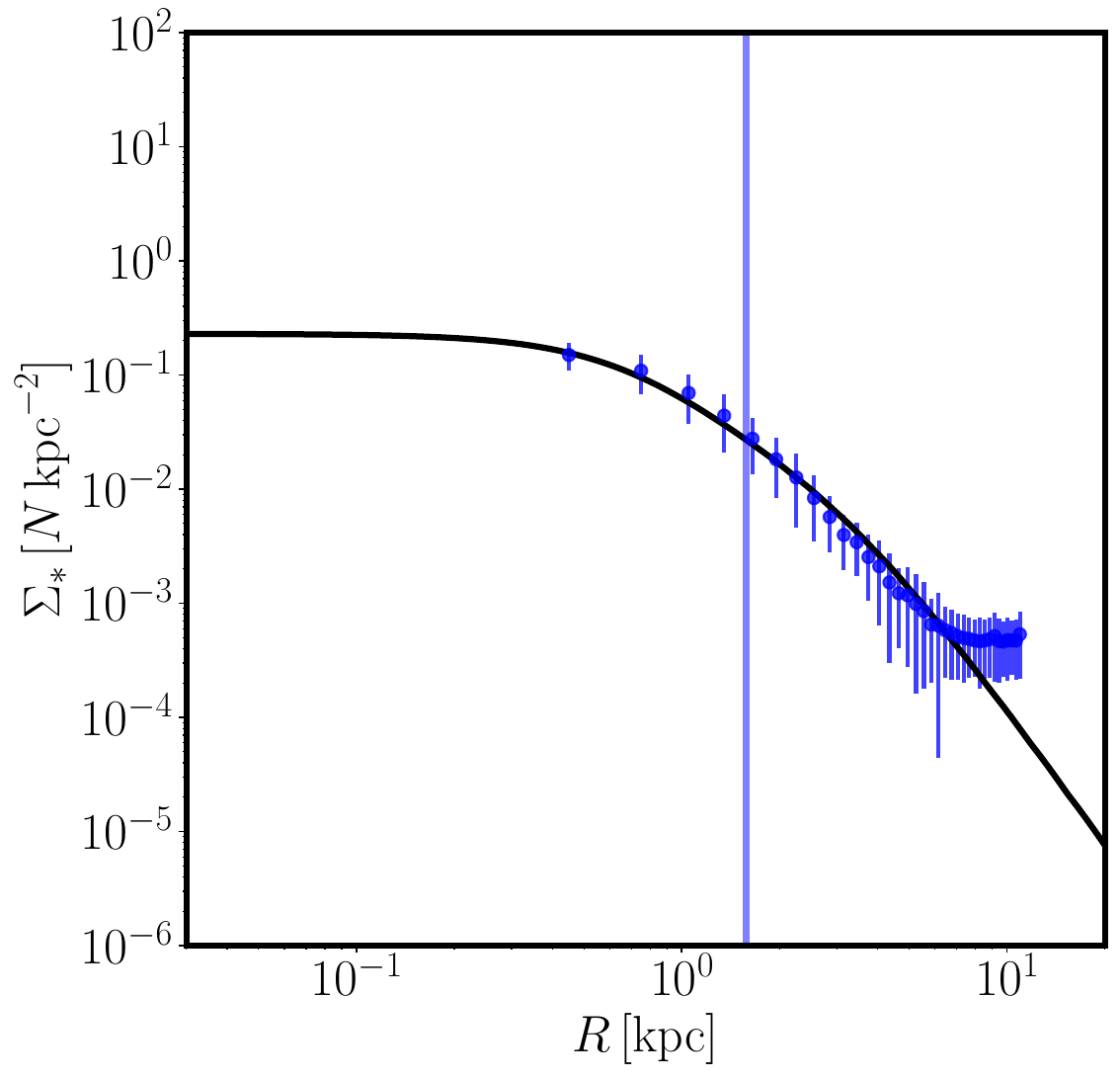}
     \end{subfigure}
     \begin{subfigure}[b]{0.45\textwidth}
         \centering
         \includegraphics[width=\textwidth, height=70mm]{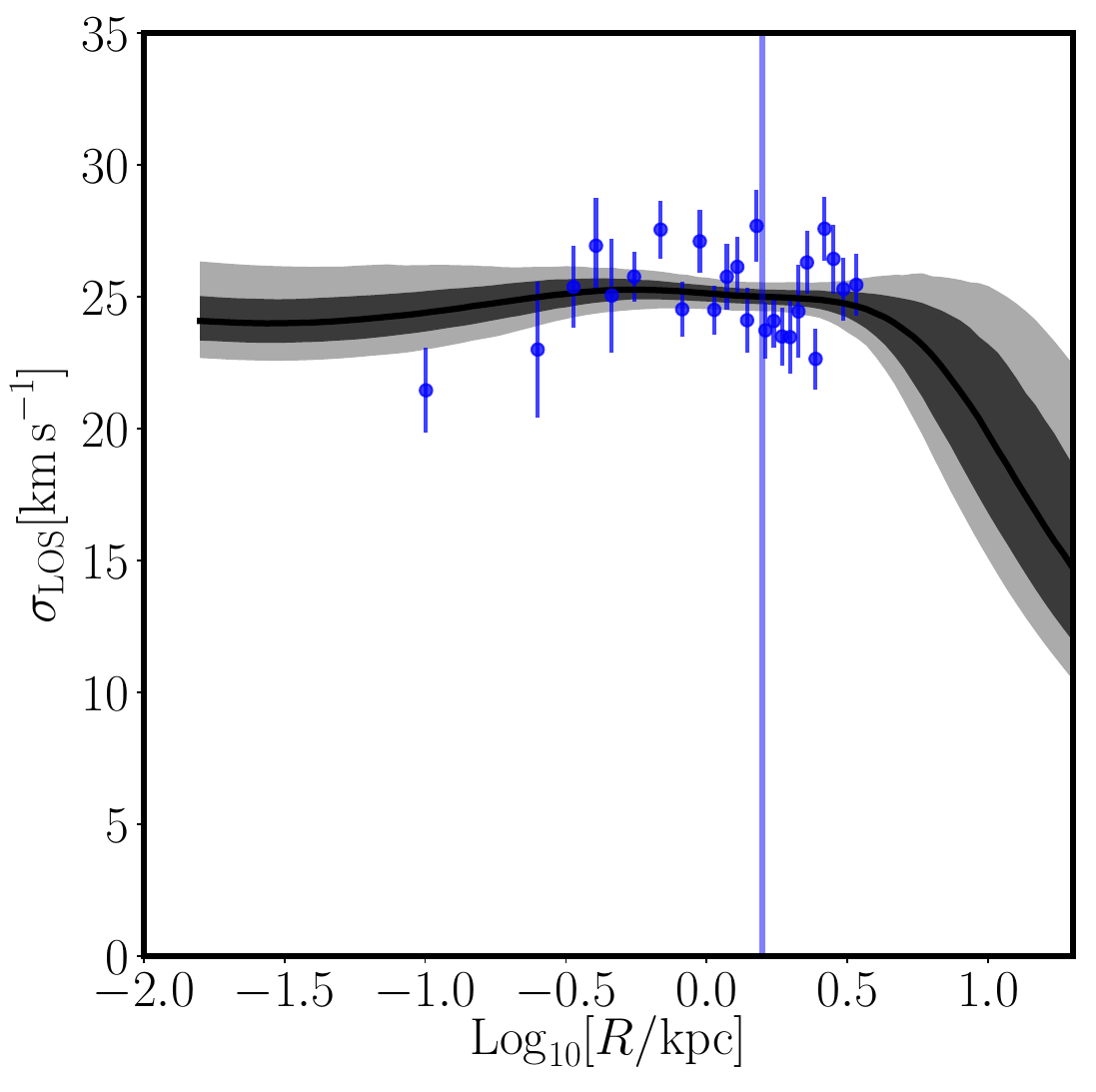}
     \end{subfigure}\\
	\centering
     \begin{subfigure}[b]{0.45\textwidth}
         \centering
         \includegraphics[width=\textwidth, height=70mm]{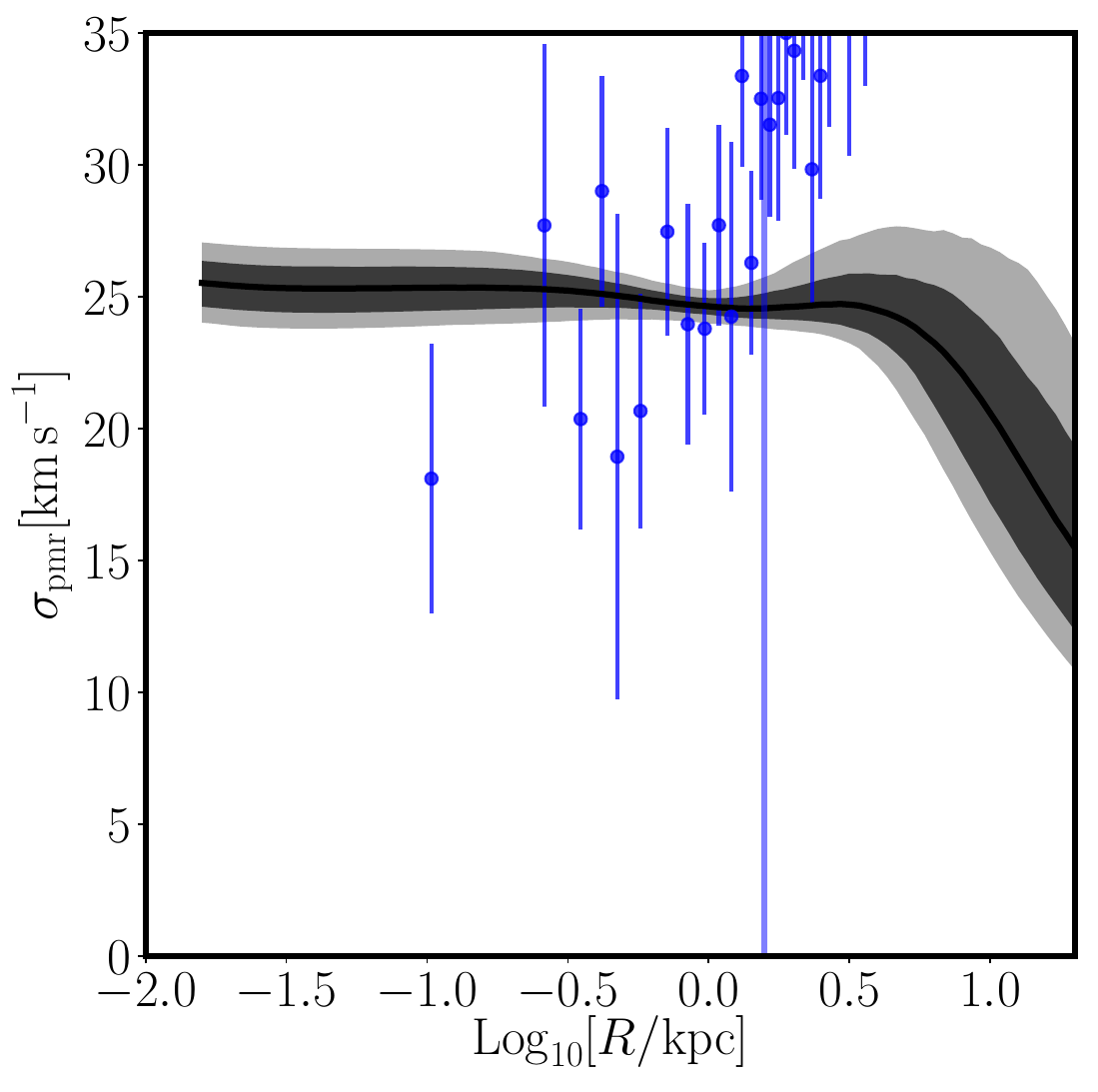}
     \end{subfigure}
     \begin{subfigure}[b]{0.45\textwidth}
         \centering
         \includegraphics[width=\textwidth, height=70mm]{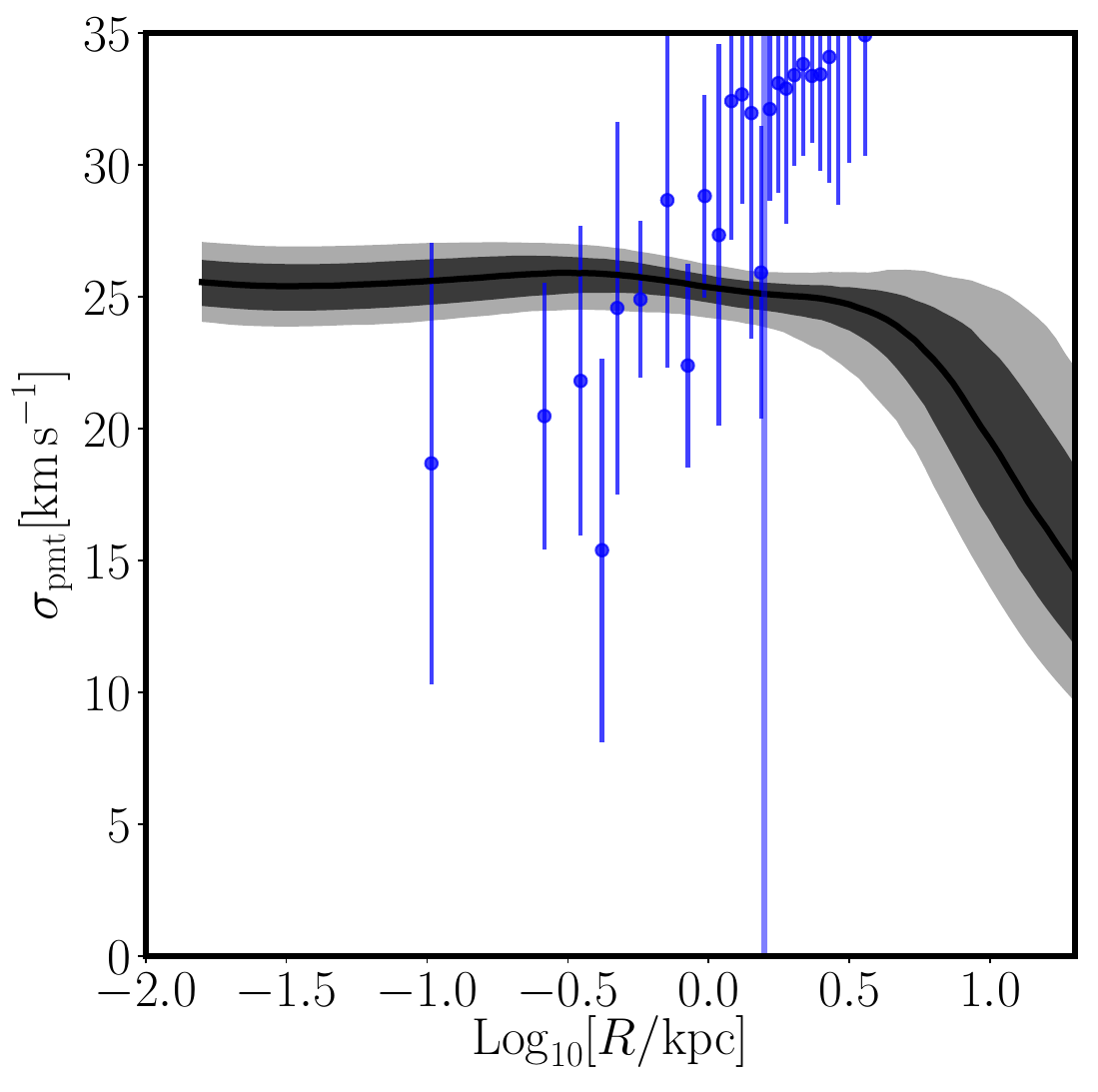}
     \end{subfigure}
        \caption{\textsc{GravSphere} recovered profiles for the real SMC. \textit{Top left panel}: surface brightness $\Sigma_*(r)$. The black line is the \textsc{GravSphere} fit, the blue points with errorbars are the observed SMASH data and the faint blue line marks the half-light radius computed by \textsc{GravSphere}. \textit{Top right panel}: velocity dispersion profile along the line of sight $\sigma_{LOS}$, the black line is \textsc{GravSphere} best fit solution, the dark and light grey contour are respectively the 68\% and the 95\% confidence intervals, the blue points with error bars are the binned data and the faint blue line marks the half-light radius computed by \textsc{GravSphere}. \textit{Bottom left panel}: velocity dispersion profile along the radial direction $\sigma_{rad}$, lines and symbols are the same as for $\sigma_{LOS}$. \textit{Bottom right panel}: velocity dispersion profile along the tangential direction $\sigma_{tan}$, lines and symbols are the same as for $\sigma_{LOS}$.}
        \label{fig:realSMC_profs}
\end{figure*}

In this Appendix, we show the surface brightness profile and the three velocity profiles (line-of-sight, radial and tangential) recovered by \textsc{GravSphere} for the case of the heavily disrupted simulation (Fig.~\ref{fig:disrupted_profs}) and for the real SMC (Fig.~\ref{fig:realSMC_profs}). As can be seen in both cases, the LOS data provides most of the constraining power to the model.

\section{Different priors}\label{app_priors}

\begin{figure*}
     \centering
     \begin{subfigure}[b]{0.48\textwidth}
         \centering
         \includegraphics[width=\textwidth, height=75mm]{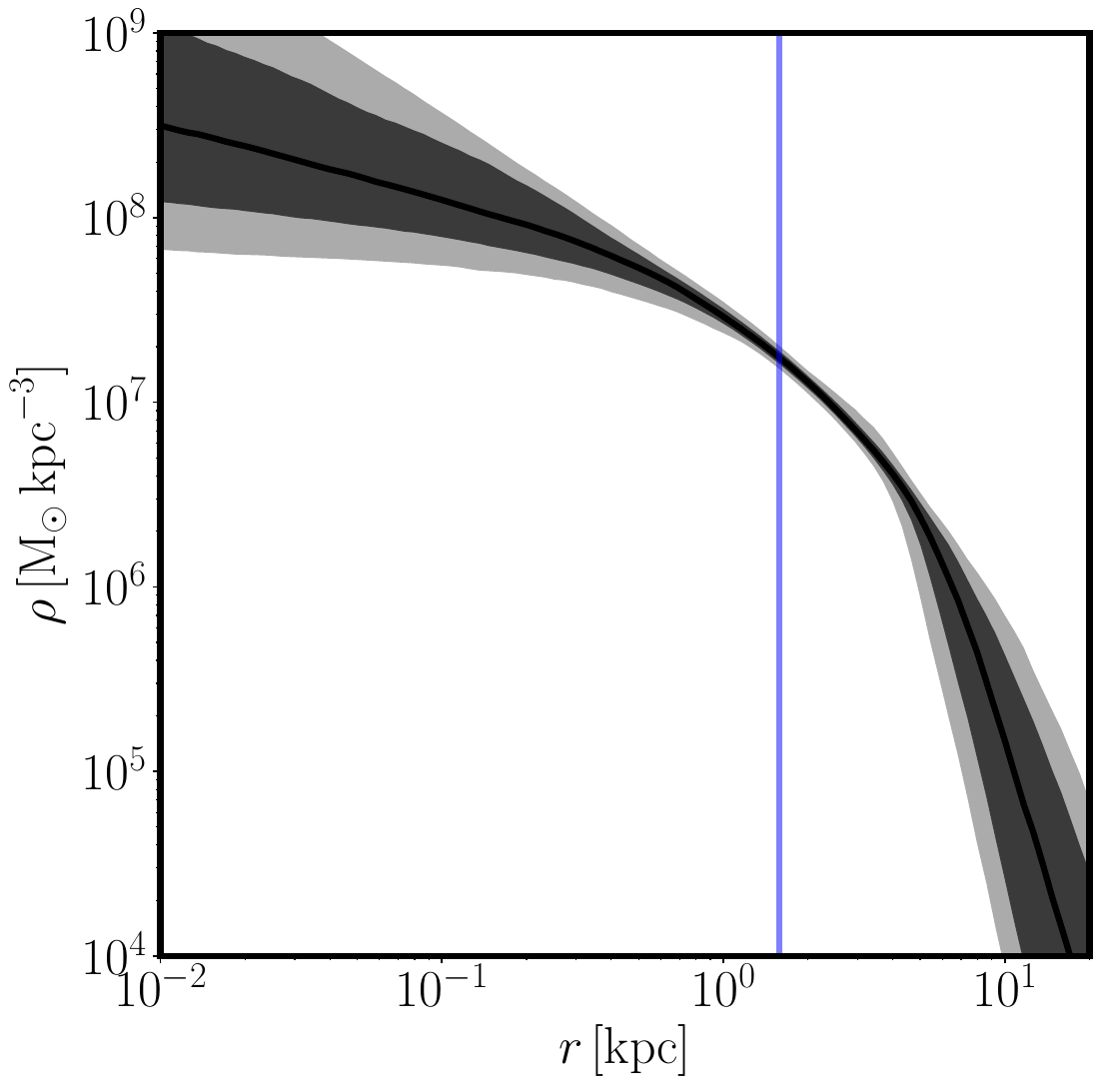}
     \end{subfigure}
     \begin{subfigure}[b]{0.48\textwidth}
         \centering
         \includegraphics[width=\textwidth, height=75mm]{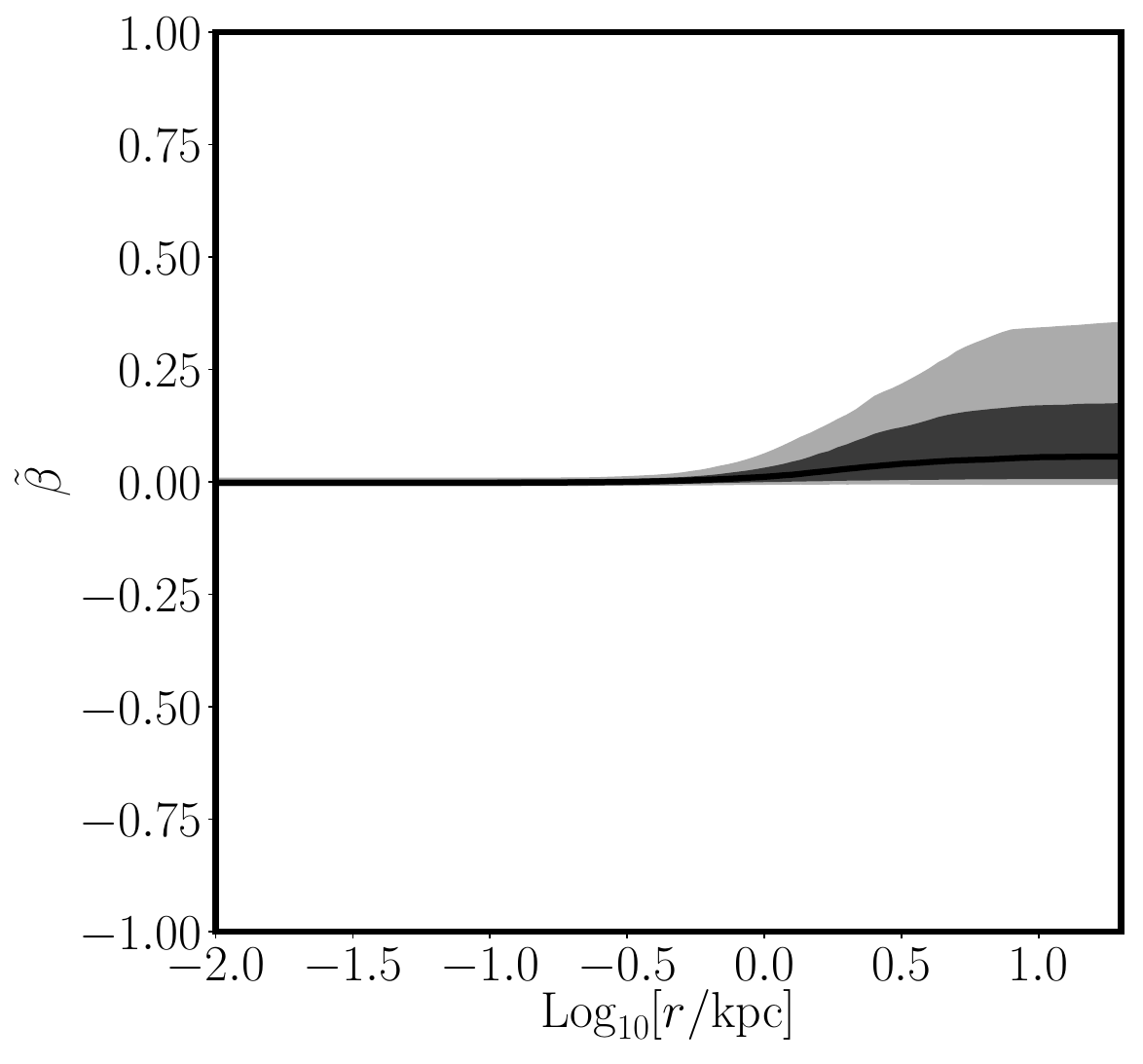}
     \end{subfigure}
        \caption{Same as Fig.~\ref{fig:smc_grav}: the mass density profile, $\rho(r)$, (\textit{left panel}) and the symmetrised anisotropy profile, $\widetilde{\beta}(r)$, (\textit{right panel}) recovered by \textsc{GravSphere} for the SMC data but for a model with prior $-0.01 \leq \beta_0 \leq 0.01$.}
       \label{fig:tight_beta_prior}
\end{figure*}

As discussed in the main text in Sec.~\ref{sec:disc_priors}, we tested the effect of different priors on our parameter recovery. Most of the tests (changing the bounds of the priors on $n$ or restricting to smaller maximum $r_c$ or $r_t$) had negligible impact on our models. The most impactful prior choice was on $\beta_0$, specifically on enforcing a tighter prior with $-0.01 \leq \beta_0 \leq 0.01$. Even this change had only a minor impact on the recovered mass density profile, $\rho(r)$, as can be seen in Fig.~\ref{fig:tight_beta_prior}. Notice that the density profile (left) now suggests a small inner core within the 95\% confidence intervals. However, the best fit results are in good agreement with our default broader priors (see Figure \ref{fig:smc_grav}) and still consistent with a dark matter cusp.

\section{Concentration parameter bias}\label{app_dicintio}

\begin{figure*}
\centering
     \begin{subfigure}[b]{0.48\textwidth}
         \centering
         \includegraphics[width=\textwidth, height=75mm]{Plots/Bouche_150_dicintio_ref_rep.pdf}
     \end{subfigure}
     \begin{subfigure}[b]{0.48\textwidth}
         \centering
         \includegraphics[width=\textwidth, height=75mm]{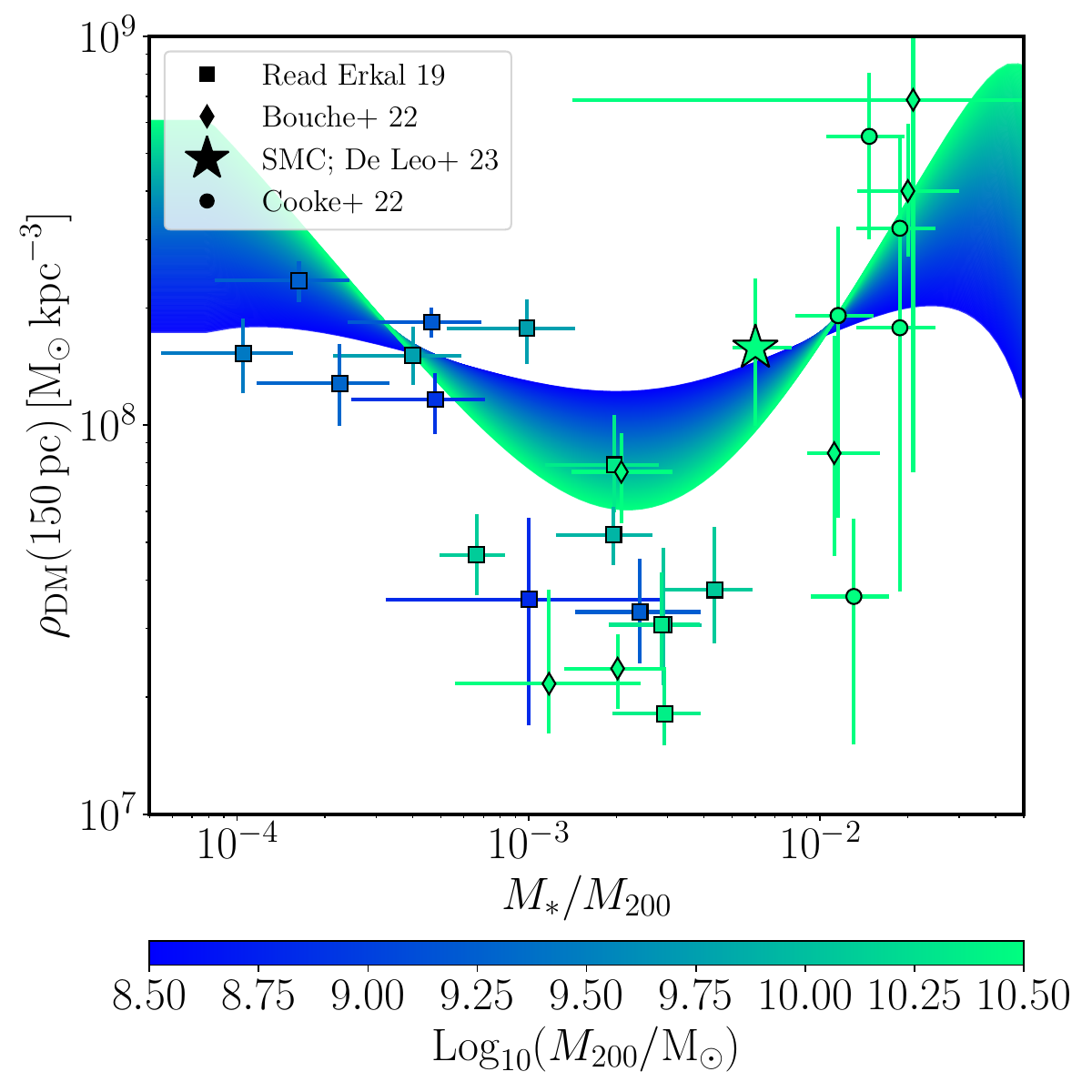}
     \end{subfigure}
\caption{Side by side comparison of the base prediction of the \citet{2014MNRAS.441.2986D} model (\textit{left panel}, same as the top right corner of Fig.~\ref{fig:multi_prof}) and the same model but factoring an hypothetical $c_{200}$ bias. Notice how the bias in the estimation of the concentration parameter can reconcile the model with some of the highest density dwarfs (including the SMC)}
\label{fig:dicintio_bias}
\end{figure*}

In this Appendix we test whether a 2-$\sigma$ bias in the estimation of the concentration parameter $c_{200}$ can reconcile the \citet{2014MNRAS.441.2986D} model with the observational data. In Fig.~\ref{fig:dicintio_bias} we show the model prediction for the median $c_{200}$ on the left panel and the model prediction for concentration parameters 2-$\sigma$ higher on the right panel. Notice how the higher density data (including the SMC) that were not fit well by the base model are now consistent with the predict curve of the right panel. This points to the fact that the discrepancy between the prediction of the \citet{2014MNRAS.441.2986D} model and the observational data has a magnitude comparable to the scatter in the $c_{200}$ estimation. In other words, this means that the data are not necessarily inconsistent with the model and the SMC and other dwarfs might be interpreted as higher concentration halos.

\bsp
\label{lastpage}
\end{document}